\documentclass[runningheads]{svmult}

\usepackage{makeidx}   % allows index generation
\usepackage{graphicx}  % standard LaTeX graphics tool
\usepackage{subeqnar}  % subnumbers individual equations
\usepackage{multicol}  % used for the two-column index
\usepackage{physprbb}  % modified textarea for proceedings,
\makeindex             % used for the subject index
\begin{document}
\title*{Calibration of the Distance Scale from Cepheids}
\toctitle{Calibration of the Distance Scale from Cepheids}
\titlerunning{Cepheid Distance Scale}
\author{Pascal Fouqu\'e\inst{1,2}
\and Jesper Storm\inst{3}
\and Wolfgang Gieren\inst{4}}
\authorrunning{Pascal Fouqu\'e et al.}
\institute{Observatoire de Paris, 
           LESIA,\\
           5, place Jules Janssen,\\
           F-92195 Meudon Cedex, France
\and European Southern Observatory,\\
     Casilla 19001,\\
     Santiago 19, Chile
\and Astrophysikalisches Institut Potsdam,\\
     An der Sternwarte 16,\\
     D-14482 Potsdam, Germany
\and Astronomy Group, Departamento de F\'{\i}sica,\\
     Facultad de Ciencias F\'{\i}sicas y Matem\'aticas,\\
     Universidad de Concepci\'on, Casilla 160-C,\\
     Concepci\'on, Chile}

\maketitle              % typesets the title of the contribution

\section{Introduction}

Since the discovery by Ms. Leavitt almost a hundred years ago that
Cepheid variables obey a tight relationship between their pulsation
periods and absolute magnitudes, astronomers have made great efforts
to calibrate this relationship, and use it to estimate the distances
to nearby galaxies in which Cepheids were found. For a very nice
review of the early history of the Cepheid period-luminosity (PL)
relation, see Fernie (\cite{f69}). With the course of time, the
calibration of the PL relation was refined, using new methods and
improving data for both Cepheids in our own Galaxy and Cepheids which
were found in increasing numbers in Local Group galaxies. In
particular, the Magellanic Clouds have played a fundamental role in
our effort to calibrate the PL relation (and still do so today, as we
will show in this review), mainly because they are just near enough to
make Cepheid apparent magnitudes bright enough for accurate photometry
with even small telescopes, and on the other hand distant enough to
have them, in a good approximation, all at the same distance. The
slope of the PL relations can thus be determined directly from a
sample of LMC Cepheids in contrast to a sample of Galactic Cepheids
where accurate individual distances, which are fundamentally difficult
to determine, are needed. In the course of the decades, it became
clear that the Cepheid PL relation is a very powerful method to
determine extragalactic distances, and it was (and still is) generally
considered as the most accurate and reliable stellar method to
calibrate the extragalactic distance scale. For that reason, the HST
Key Project on the Extragalactic Distance Scale chose the strategy to
detect samples of Cepheids in a number of selected late-type galaxies
and use them to measure the distances to these galaxies, which then
served to calibrate other, more far-reaching methods of distance
measurement to determine the Hubble constant in a region of constant
Hubble flow. It is clear that we have gone a very long way from the
early attempts to calibrate the PL relation, to the application of
this technique to Cepheids in stellar systems as distant as 20 Mpc, as
successfully done by the groups who have used the Hubble Space
Telescope for this purpose.

In spite of all these successes, it has also become clear over the
past decade that there are still a number of problems with the
calibration of the PL relation which so far have prevented truly
accurate distance determinations, to 5 percent or better, as needed
for the cosmological, and many other astrophysical applications.  One
basic problem has been the notorious difficulty to measure accurate,
independent distances to Galactic Cepheids needed for a calibration of
the PL relation in our own Galaxy (see next section). The alternative
approach, used many times, is to calibrate the PL relation in the LMC,
but this requires an independent knowledge of the LMC distance whose
determination has proven to be amazingly difficult (see the review of
A.R. Walker in this volume). Another problem complicating the
calibration of the PL relation is that Cepheids, as young stars, tend
to lie in crowded and dusty regions in their host galaxies, making
absorption corrections a critical issue.  In more recent years, work
on the PL relation has therefore increasingly shifted to the
near-infrared where the problems with reddening are strongly reduced
as compared to the optical spectral region. Another potential problem
with the use of the Cepheid PL relation is its possible sensitivity to
chemical abundances; if such a metallicity dependence exists and is
significant, it has to be taken into account when comparing Cepheid
populations in different galaxies which have different
metallicities. Therefore, while there has been a lot of progress on
the calibration of Cepheids as distance indicators over the years,
there is still room (and need) for a substantial improvement. It is
the purpose of this review to contribute such progress, and our approach
is to combine Galactic and LMC Cepheids in the best possible way to
derive both an improved absolute calibration of the PL relation in a
number of optical and near-infrared photometric bands and, in a
parallel step, derive an improved distance to the Large Magellanic
Cloud from its Cepheid variables. One of the reasons why a PL
calibration from Galactic Cepheid variables is of advantage as
compared to a calibration based on LMC Cepheids alone is the fact that
in most large spiral galaxies, and in particular in those targeted by
the HST Key Project, the mean metallicities are quite close to solar,
implying that metallicity-related systematic effects are minimized
when comparing these extragalactic PL relations to the Galactic one,
rather than to the one defined by the more metal-poor population of
LMC Cepheids.

In a final step, we will test what our new Cepheid PL calibration
implies for other stellar candles frequently used for distance work,
such as RR Lyrae stars, red giant clump stars, and the tip of the red
giant branch.

\section{The Galactic vs. the LMC routes}

\subsection{The infrared surface brightness method}

Fifteen years ago, the classical method of calibrating the PL relation
for Cepheids was to use the ZAMS-fitting technique to determine the
distances of a handful of open clusters which happen to contain
Cepheid members (Feast \& Walker \cite{fw87}). However, Hipparcos
revealed that the distance of the calibrating cluster, the Pleiades,
had to be revised substantially downwards, at a level where the
distance difference between Hyades and Pleiades can no longer be explained
only by metallicity differences. Therefore, some doubts were shed on
the ZAMS-fitting technique, and it became necessary to find
alternative techniques of similar accuracy. It is a measure of our
progress to see that two such methods have emerged in the meantime.

The main alternative method, based on the classical ideas of Baade and
Wesselink, and first implemented by Barnes \& Evans (\cite{be76})
consists in combining linear diameter measurements, as obtained from
radial velocity curve integration, to angular diameter determinations
coming from measurements of magnitudes and surface brightness to
derive the mean diameter and distance of a Cepheid. The surface
brightness estimates come from a relation between this parameter and a
suitably chosen colour.

In a comparison of the results of both methods, Gieren \& Fouqu\'e
(\cite{gf93}) established that the Barnes-Evans zero point of the
PL relation in the $V$ band was 0.15 mag brighter than the ZAMS-fitting
zero point. However, the Barnes-Evans method uses the $V-R$ colour index
to estimate the surface brightness, and it was soon discovered that a
much better estimate could come from infrared colours (Welch
\cite{w94}; Laney \& Stobie \cite{ls95}).

Encouraged by the very promising near-infrared results, Fouqu\'e \&
Gieren (\cite{fg97}) calibrated the infrared surface brightness
technique, using both $J-K$ and $V-K$ colours, by assuming that
non-variable, stable giants and supergiants follow the same surface
brightness vs. colour relation as the pulsating Cepheids. Using 23
stars with measured angular diameters, mostly from Michelson
interferometry, they checked that the slope of the relation directly
derived from Cepheids was consistent, within very small uncertainties,
with the slope derived from stable stars. This provided confidence to
also adopt the zero point from the giants and supergiants. They
recalibrated the Barnes-Evans relation and showed that the accuracy of
the infrared method for deriving the distances and radii of individual
Cepheids was 5 to 10 times better than the results produced by an
application of the optical counterpart of the technique. As in the
optical surface brightness technique, a very important feature and
advantage of the infrared surface brightness method is its very low,
and almost negligible dependence on absorption corrections.

At that time, only one Cepheid ($\zeta$ Gem) angular diameter had been
measured, with the lunar occultation technique (Ridgway et
al. \cite{r+82}) and the agreement with our predicted angular diameter
led some support to our choice of the zero point. However, that
comparison suffered from the relatively large error of the Cepheid
angular diameter measurement.

More recently, Nordgren et al. (\cite{n+02}) have confirmed our
calibrating surface brightness-colour relations from an enlarged
sample of 57 giants with accurate interferometric measurements of
their angular diameters. Interestingly, they find a similar scatter to
ours in these relations, which probably means that intrinsic
dispersion has been reached. Then, they used 59 direct interferometric
diameter measurements for 3 Cepheids to compute their surface
brightnesses, at the corresponding pulsation phases. From these
measurements, they derived surface brightness-colour relations for the
first time directly from the Cepheids themselves, and confirmed that
the Cepheid surface brightnesses do indeed follow the calibrating
relations obtained from stable giants and supergiants in the same
colour range as Cepheids, yielding a zero point fully compatible with
our previous value from stable stars ($3.941 \pm 0.004$ vs. $3.947 \pm
0.003$, respectively). Subsequently, Lane et al. (\cite{la+02}) were
able to go a step further and measure the angular diameter variations
for 2 Cepheids, therefore allowing a measure of their distances and
mean diameters independently of photometric measurements, but also
confirming the adopted calibrating relations.

At the time of this review, three Cepheids have distance
determinations based on interferometric measurements of their angular
diameters. It is instructive to compare them to the distances derived
from trigonometric parallaxes. This is done in Table~\ref{Tab1}. In
the case of $\delta$ Cep, we have used the recent HST measurement by
Benedict et al. (\cite{b+02}), which supersedes the less accurate
Hipparcos measurement. For $\zeta$ Gem, the trigonometric parallax
comes from Hipparcos, while for $\eta$ Aql we have used a weighted
mean of Hipparcos and USNO measurements, as in Nordgren et
al. (\cite{n+00}). The agreement is very good, especially in the case
of the accurate trigonometric measurement of $\delta$ Cep. Note that
the small uncertainty associated with the interferometric distance
determination of $\delta$ Cep neglects the possible systematic
uncertainties introduced by the use of the surface brightness
vs. colour relations.

\begin{table}
\caption{Comparison of Cepheid distances from interferometry and trigonometric parallaxes}
\begin{center}
\renewcommand{\arraystretch}{1.4}
\setlength\tabcolsep{5pt}
\begin{tabular}{lll}
\hline\noalign{\smallskip}
Cepheid & $d_{\rm interferometry}$ & $d_{\rm trigonometry}$\\
\noalign{\smallskip}
\hline
\noalign{\smallskip}
$\delta$ Cep & $272 \pm 6$  & $273^{\;+12}_{\;-11}$  \\
$\eta$ Aql   & $320 \pm 32$ & $382^{\;+150}_{\;-84}$ \\
$\zeta$ Gem  & $362 \pm 38$ & $358^{\;+147}_{\;-81}$ \\
\hline
\end{tabular}
\end{center}
\label{Tab1}
\end{table}

\begin{figure}[ht]
\begin{center}
\includegraphics[width=.8\textwidth]{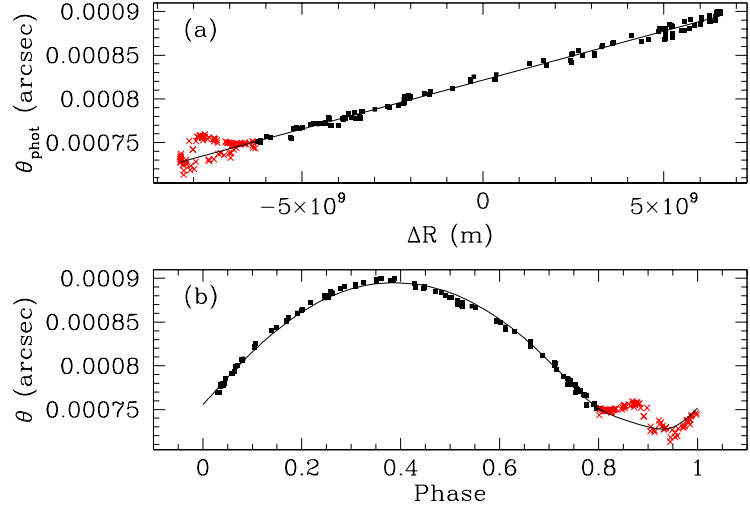}
\end{center}
\caption[]{Illustration of the ISB method in the case of X Cyg: the
  points represent the photometrically determined angular diameters,
  and the line in panel (a) shows the bisector fit to the filled
  points. The curve in panel (b) delineates the angular diameter
  obtained from integrating the radial velocity curve at the derived
  distance. Red crosses in both panels represent points which were
  eliminated before the fit. This is necessary because near minimum
  radius the existence of shock waves in the Cepheid atmosphere, and
  possibly other effects, do not allow a reliable calculation of the
  angular diameter from the photometry}
\label{Fig1}
\end{figure}

\begin{figure}[ht]
\begin{center}
\includegraphics[width=.4\textwidth]{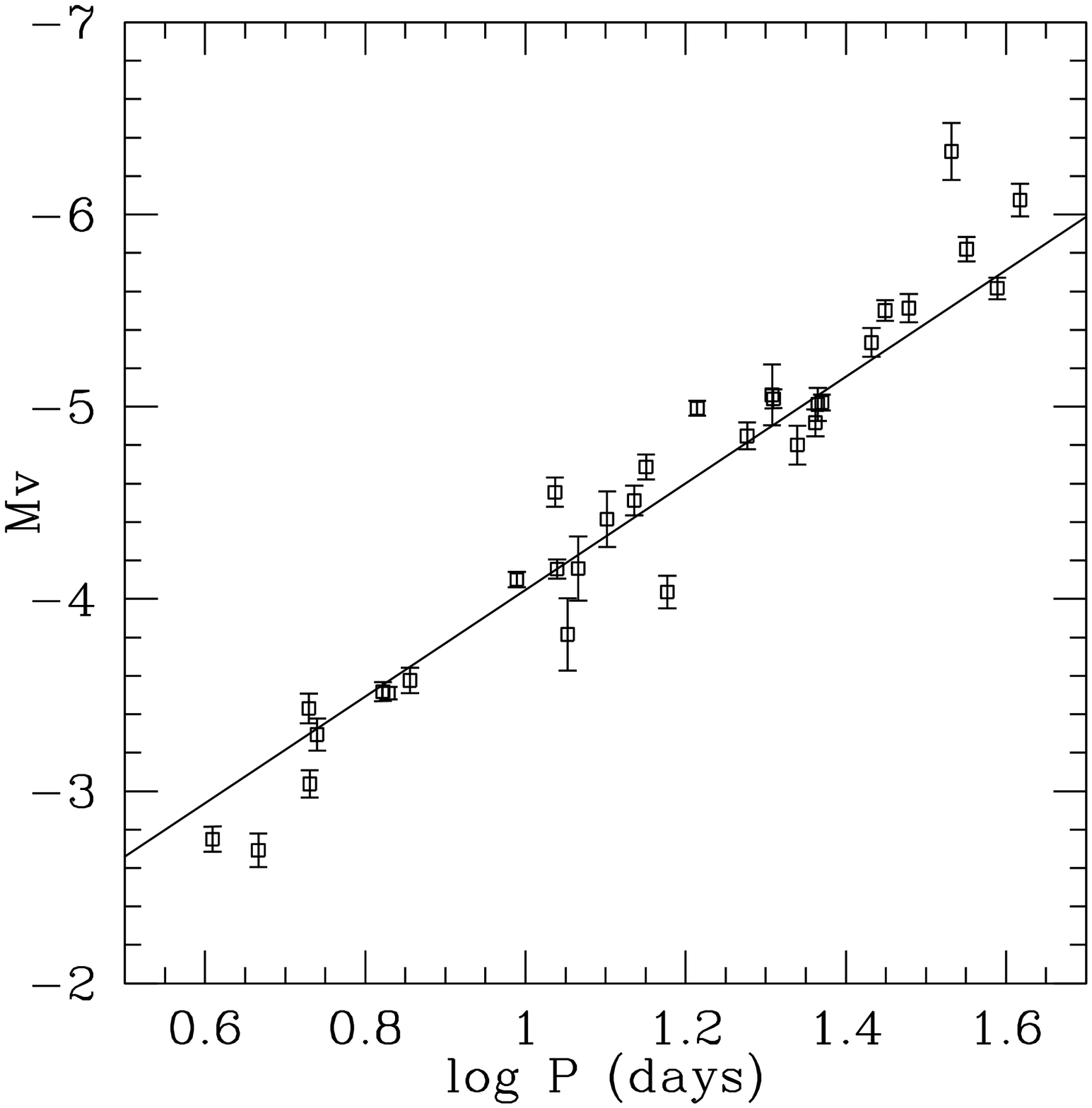}
\includegraphics[width=.4\textwidth]{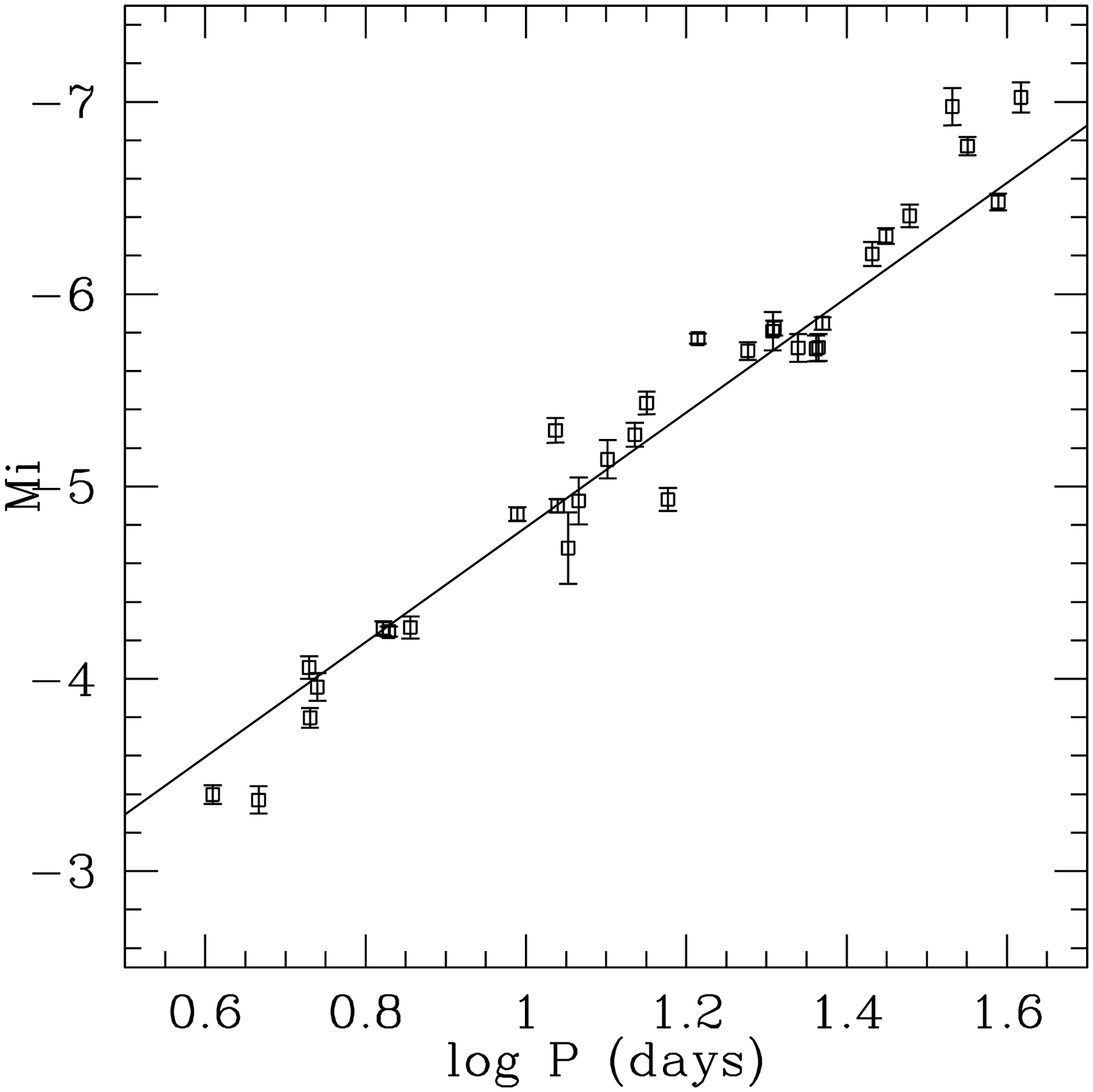}
\includegraphics[width=.4\textwidth]{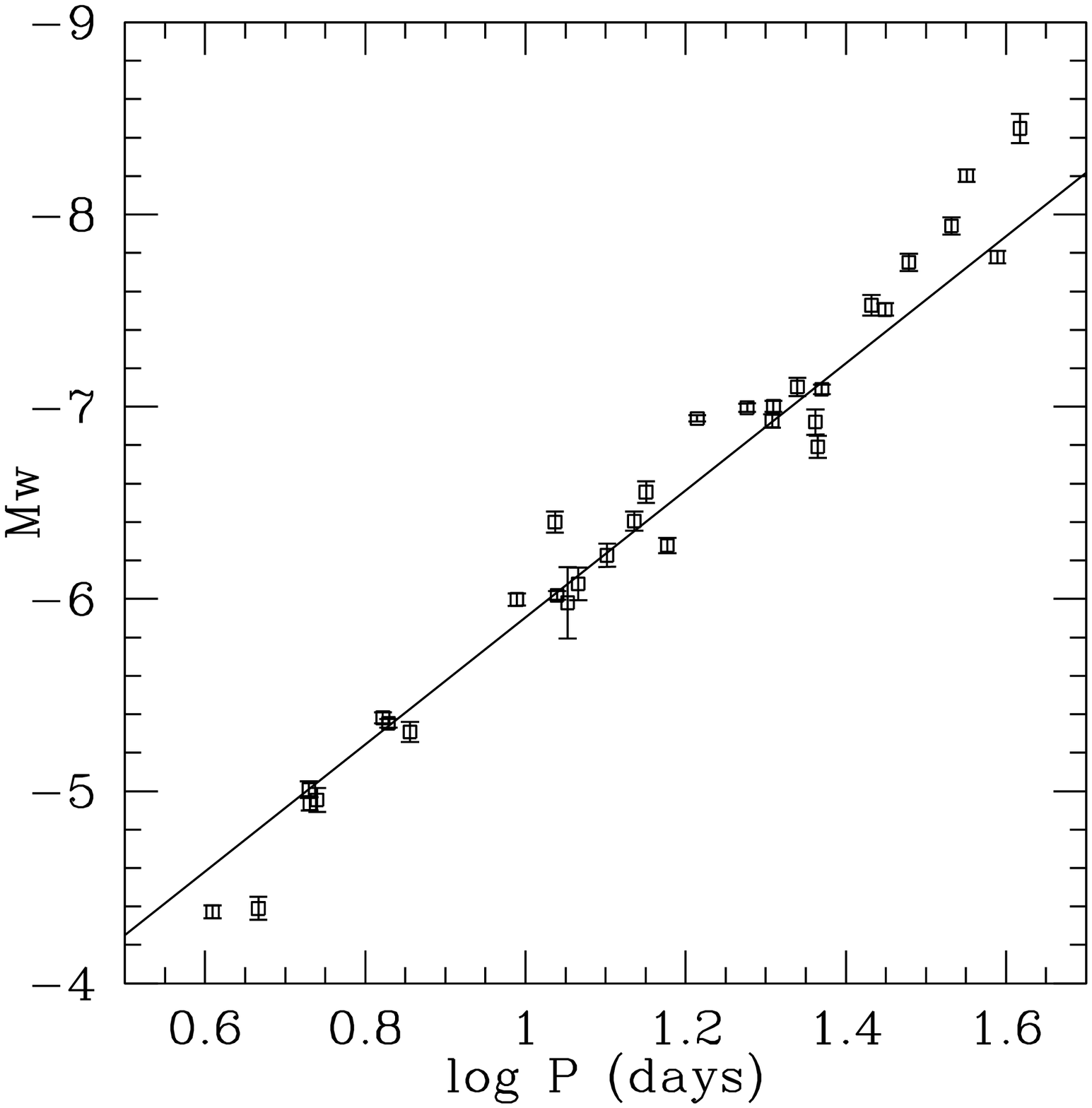}
\includegraphics[width=.4\textwidth]{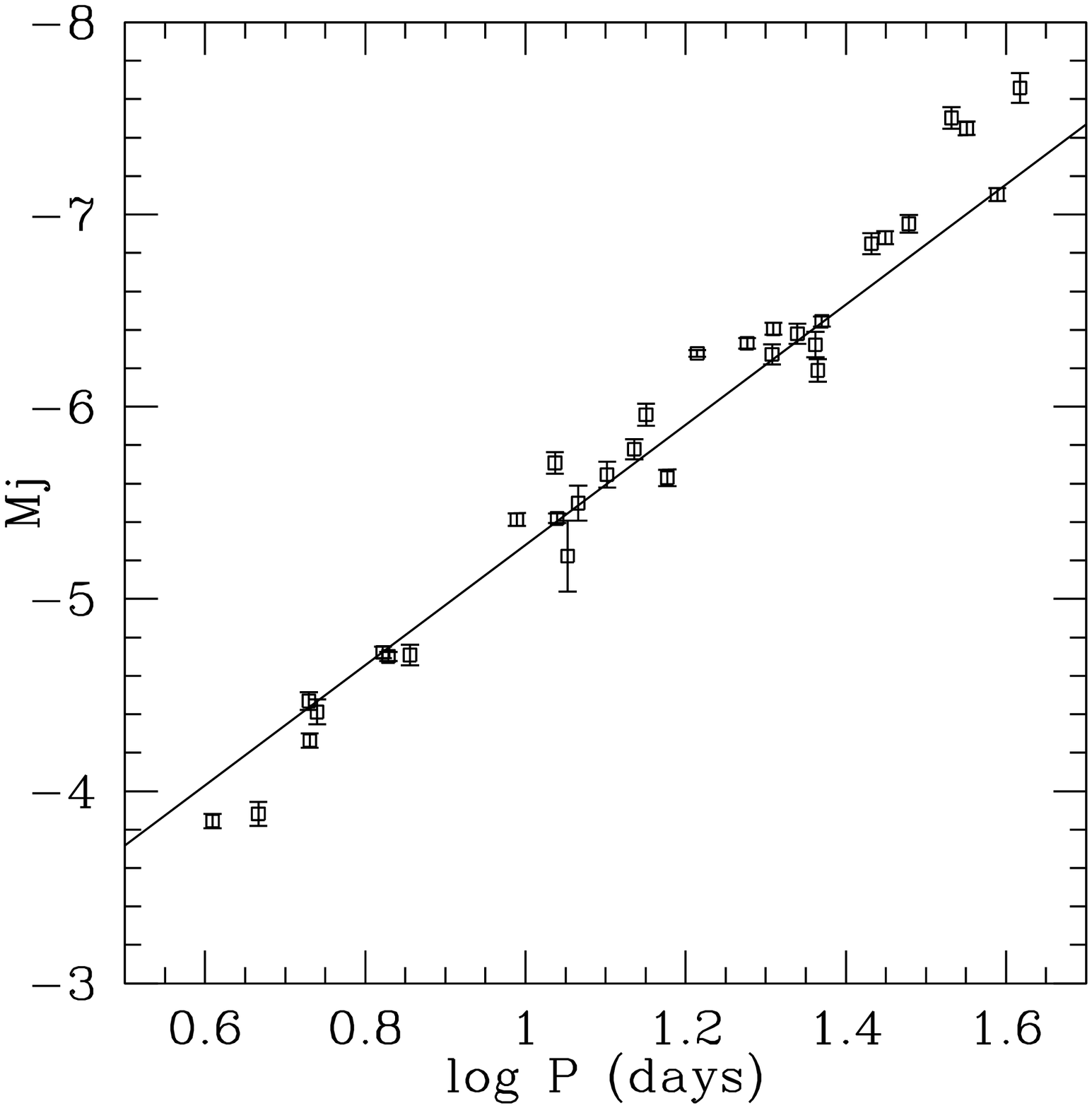}
\includegraphics[width=.4\textwidth]{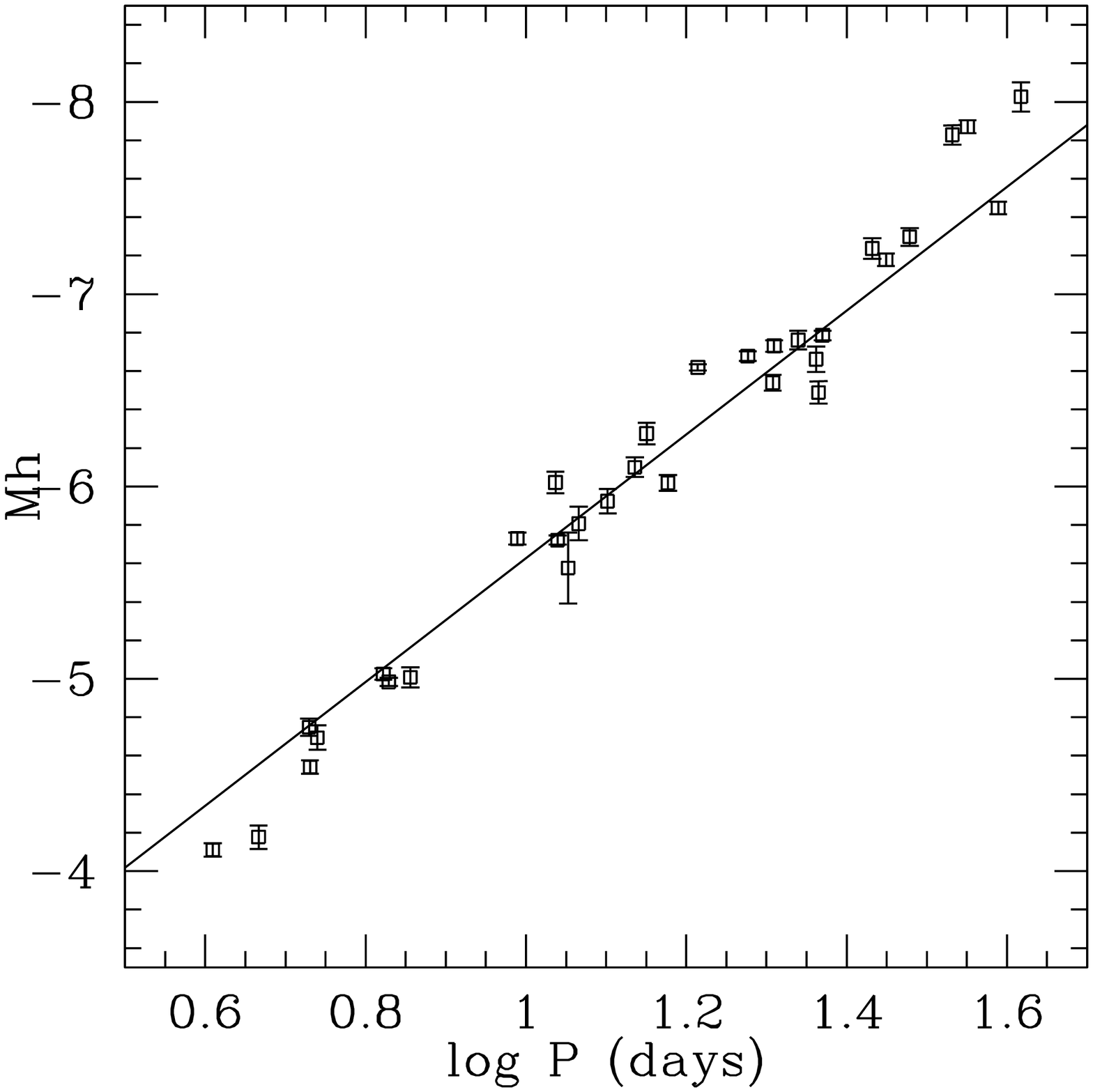}
\includegraphics[width=.4\textwidth]{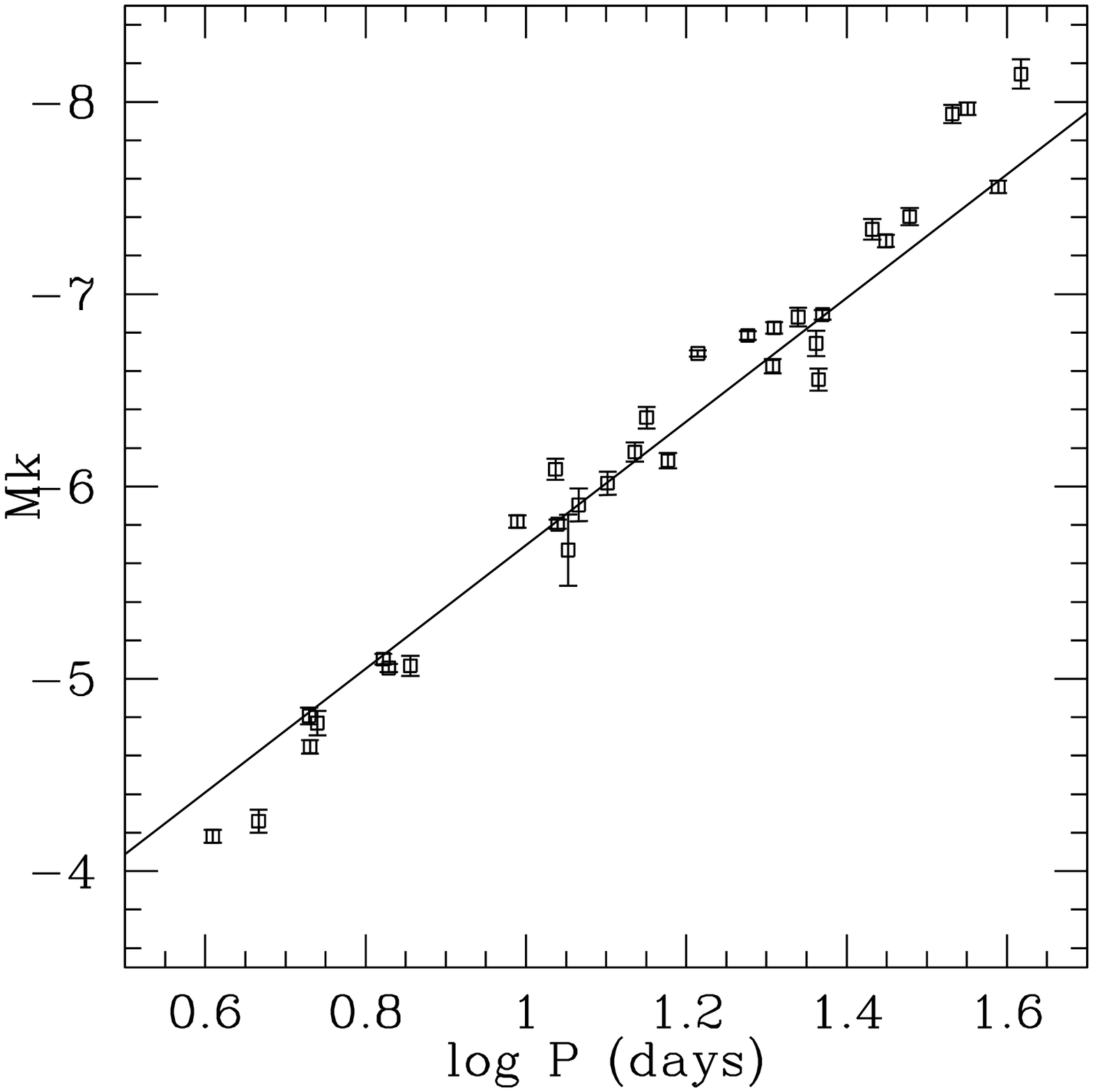}
\end{center}
\caption[]{Galactic PL relations in $VIWJHK$ bands, determined from
our new infrared surface brightness distance solutions for 32 Galactic
Cepheid variables. Superimposed lines correspond to the LMC PL
relations from OGLE2 data, adopting $\mu (\rm LMC) = 18.50$ and a mean
$E(B-V) = 0.10$}
\label{Fig2}
\end{figure}

Using the Fouqu\'e \& Gieren (\cite{fg97}) calibration, Gieren et
al. (\cite{gfg98}) derived a new calibration of the PL relation in
$VIJHK$ bands, based on 28 Galactic Cepheids with distances determined
from the infrared surface brightness method. However, determining the
slope of a linear relation from only 28 points is not very accurate,
so they chose to fix the slopes to the better-determined values from
LMC Cepheid samples, implicitly assuming that there is no metallicity
dependence of the slopes, at least in the metallicity range bracketed
by these two galaxies. More recently, we have revised the calibrating
sample to 32 Galactic Cepheids (Storm et al. \cite{s+02}), using a number of
additional Cepheid variables not used in our previous studies, and
also using fresh data from the literature whenever they had become
available. The new Cepheid distance solutions from the infrared
surface brightness technique are presented in Table~\ref{Tab7}.
Reddenings were adopted from Fernie's database (\cite{f+95}, column
labelled FE1). In Fig.~\ref{Fig1}, we show one such solution for the
Cepheid X Cyg which is fairly representative for our whole, updated
sample of Galactic Cepheid variables. Our new Galactic Cepheid
distance data confirm that the Galactic slopes of the PL relation are
steeper than their LMC counterparts, in all photometric bands, as can
be seen in Table~\ref{Tab2}. The corresponding Galactic Cepheid PL
relations are shown in Fig.~\ref{Fig2}.

\begin{table}
\caption{Slopes of various PL relations in $BVIWJHK$ bands (see
  explanations in the text)}
\begin{center}
\renewcommand{\arraystretch}{1.4}
\setlength\tabcolsep{2pt}
\begin{tabular}{ccccc}
\hline\noalign{\smallskip}
Band & Galactic slopes (N) & \multicolumn{3} {c} {LMC slopes} \\ \cline{3-5} & & literature (N) & revised (N) & E(B-V)=0.10 \\
\noalign{\smallskip}
\hline
\noalign{\smallskip}
$B$ & $-2.72 \pm 0.12$ (32) &                          &                          &                    \\
$V$ & $-3.06 \pm 0.11$ (32) & $-2.775 \pm 0.031$ (651) & $-2.735 \pm 0.038$ (644) & $-2.774 \pm 0.042$ \\
$I$ & $-3.24 \pm 0.11$ (32) & $-2.977 \pm 0.021$ (661) & $-2.962 \pm 0.025$ (644) & $-2.986 \pm 0.027$ \\
$W$ & $-3.57 \pm 0.10$ (32) & $-3.300 \pm 0.011$ (668) & $-3.306 \pm 0.013$ (644) & $-3.306 \pm 0.013$ \\
$J$ & $-3.53 \pm 0.09$ (32) & $-3.144 \pm 0.035$ (490) & $-3.112 \pm 0.036$ (447) & $-3.127 \pm 0.036$ \\
$H$ & $-3.64 \pm 0.10$ (32) & $-3.236 \pm 0.033$ (493) & $-3.208 \pm 0.034$ (447) & $-3.216 \pm 0.034$ \\
$K$ & $-3.67 \pm 0.10$ (32) & $-3.246 \pm 0.036$ (472) & $-3.209 \pm 0.036$ (447) & $-3.215 \pm 0.037$ \\
\hline
\end{tabular}
\end{center}
\label{Tab2}
\end{table}

However, there are systematic differences in the way the slopes given
in Table~\ref{Tab2} have been determined.  For instance, the reddening
corrections do not follow exactly the same law in Gieren et
al. (\cite{gfg98}) and in the work of the OGLE team (hereafter OGLE2),
and Groenewegen (\cite{g00}). Even the definition of the
reddening-free parameter $W$ varies in the literature. In order to make
things fully comparable, we have derived new LMC Cepheid PL relations
in the optical ($VIW$) from the published OGLE2 database, and in the
infrared ($JHK$) from the sample kindly provided by M. Groenewegen,
adopting the same reddening law as for our Galactic
calibrators. For this, we have computed the values of the various
coefficients $R_v$, $R_i$, $R_w$, $R_j$, $R_h$, $R_k$ for each
calibrator according to the following formulae (from Laney \& Stobie
\cite{ls93} and Caldwell \& Coulson \cite{cc87}):

\begin{eqnarray}
R_v & = & \frac{A_v}{E(B-V)} = 3.07 + 0.28 \times (B-V)_{\circ} + 0.04 \times E(B-V) \\
R_i & = & \frac{A_i}{E(B-V)} = 1.82 + 0.205 \times (B-V)_{\circ} + 0.0225 \times E(B-V) \\
R_w & = & \frac{1}{1-R_i/R_v} \\
R_j & = & \frac{A_j}{E(B-V)} = R_v/4.02 \\
R_h & = & \frac{A_h}{E(B-V)} = R_v/6.82 \\
R_k & = & \frac{A_k}{E(B-V)} = R_v/11
\end{eqnarray}

$R_w$ defines the Wesenheit magnitude as:

\begin{equation}
W = V - R_w \  (V-I)
\end{equation}

Then, we have computed the mean value of these coefficients over our
32 Galactic calibrators, assumed to be representative for the entire
Galactic Cepheid population. As the rms dispersions turned out to be
small (from 0.003 in $K$ to 0.036 in $V$), we decided to adopt the same
constant values for all the Cepheids. These are:

\begin{eqnarray}
R_v & = & 3.30 \\
R_i & = & 1.99 \\
R_w & = & 2.51 \\
R_j & = & 0.82 \\
R_h & = & 0.48 \\
R_k & = & 0.30
\end{eqnarray}

\begin{figure}[ht]
\begin{center}
\includegraphics[width=.4\textwidth]{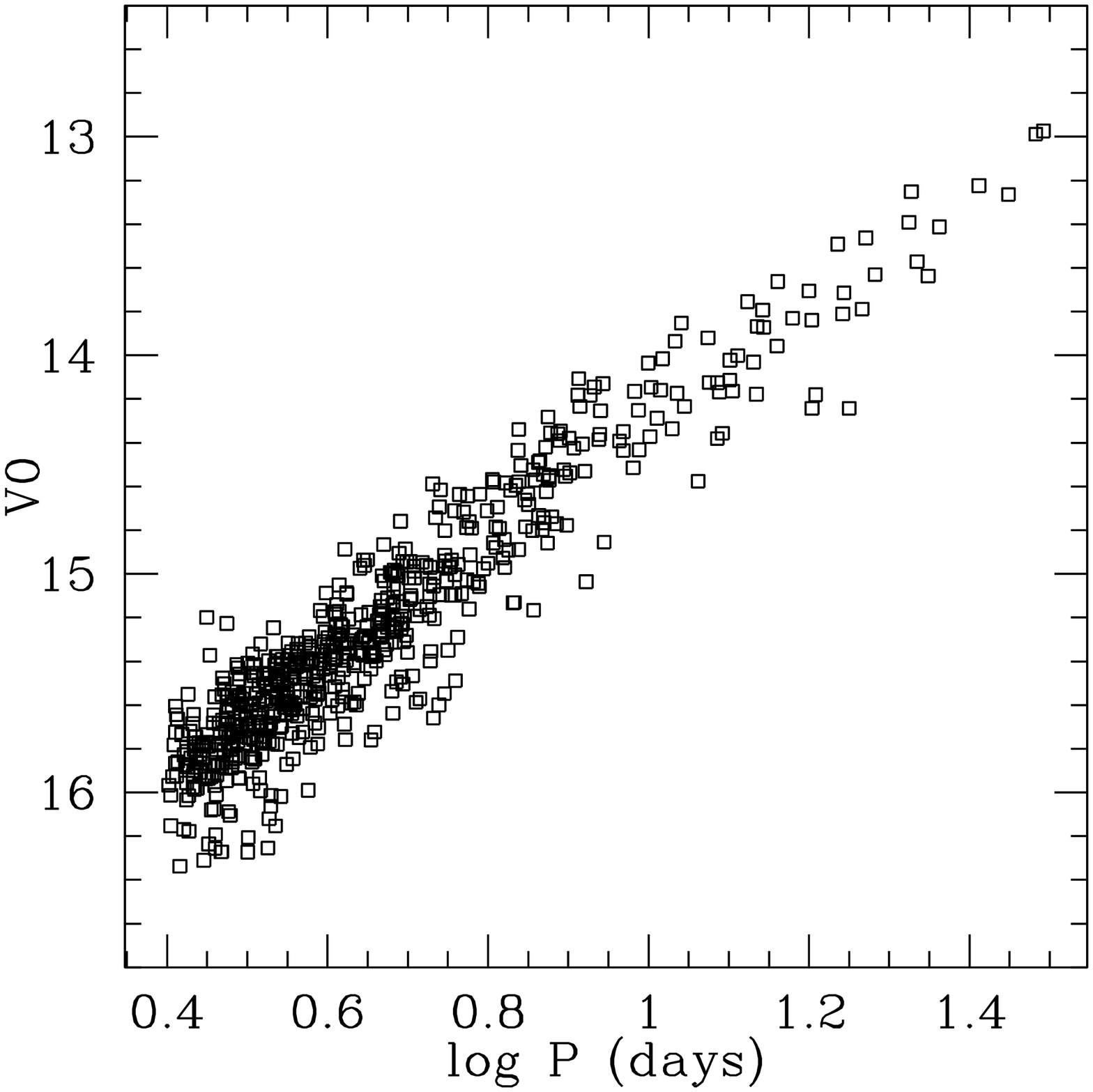}
\includegraphics[width=.4\textwidth]{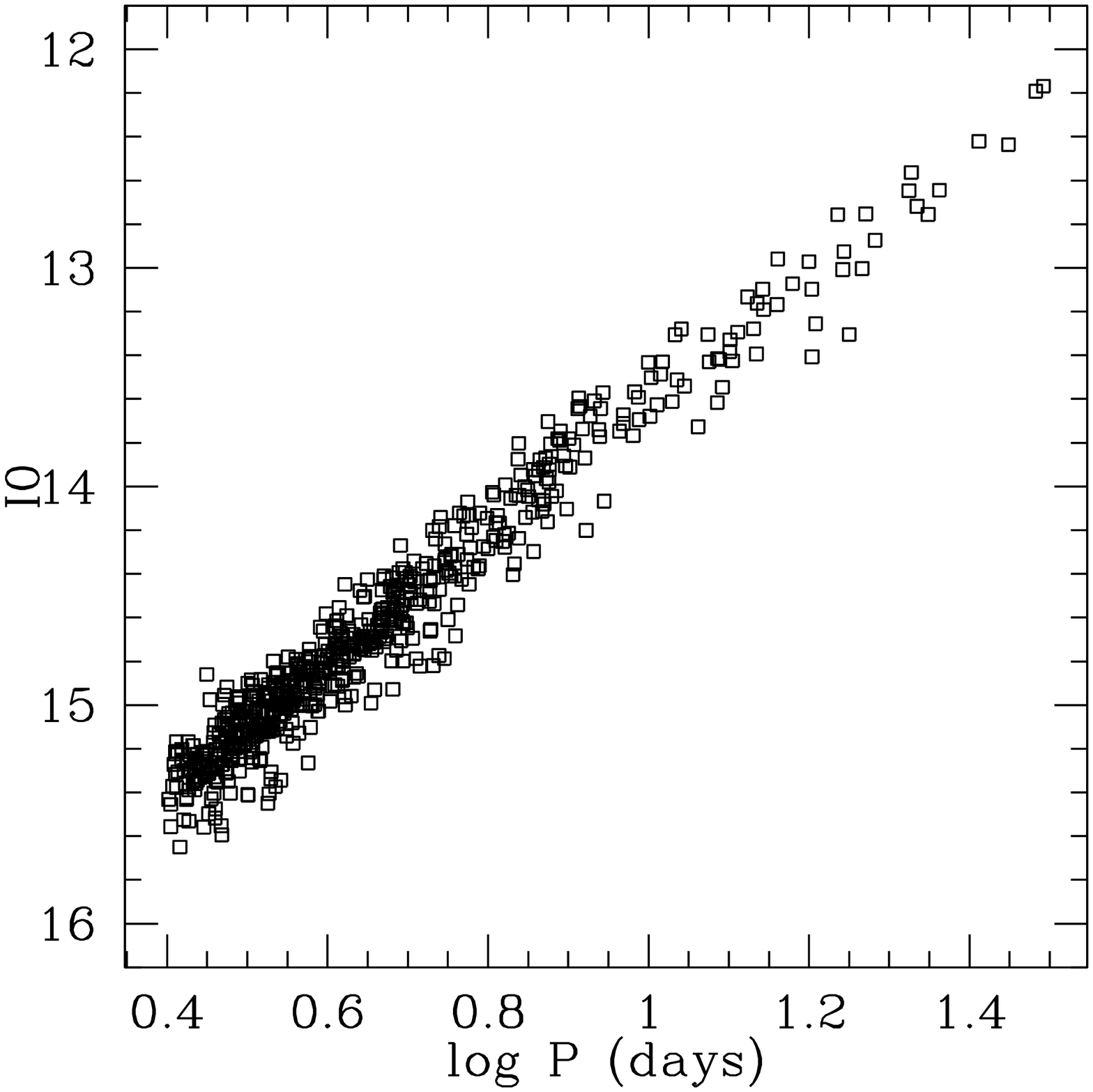}
\includegraphics[width=.4\textwidth]{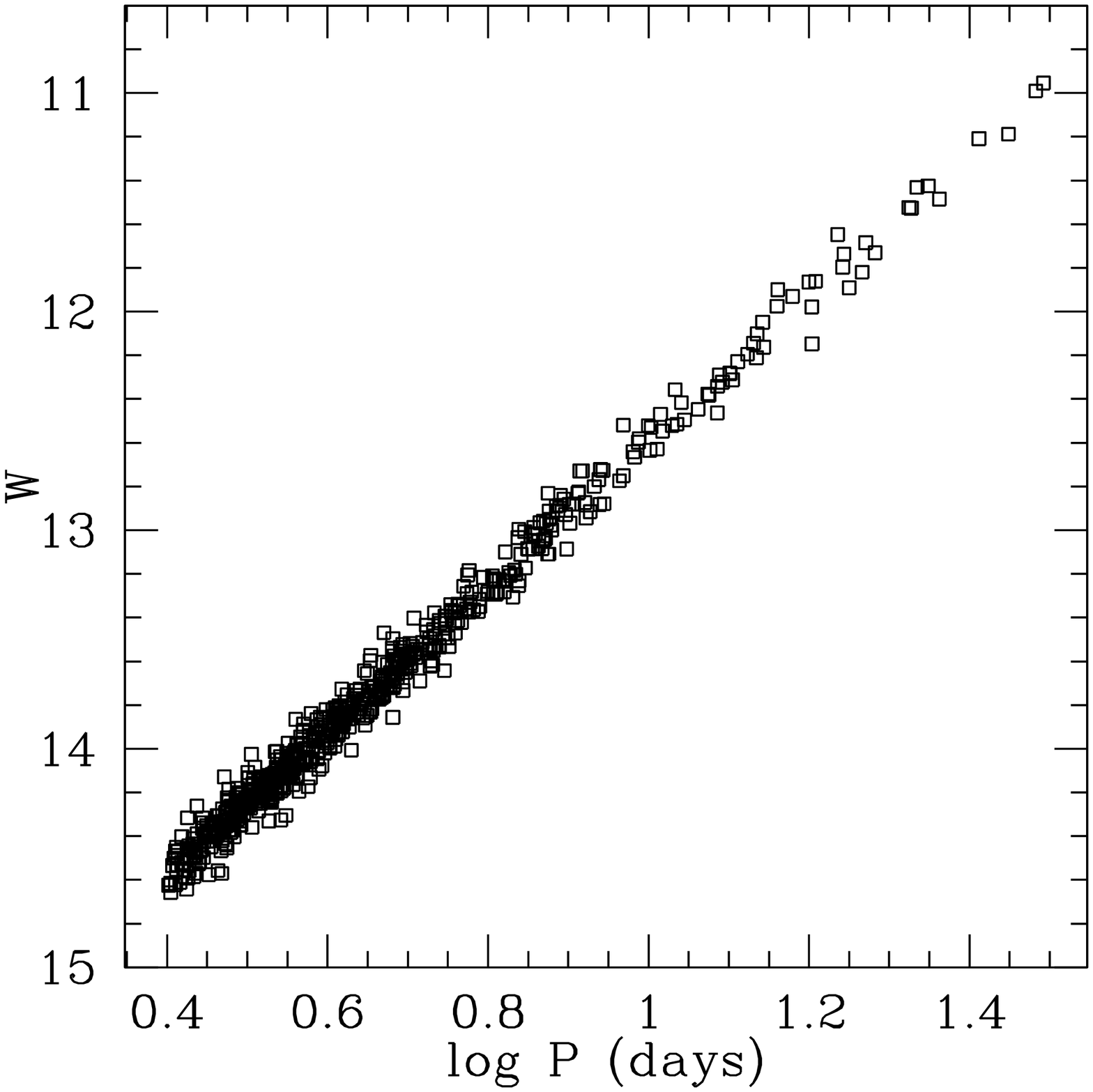}
\includegraphics[width=.4\textwidth]{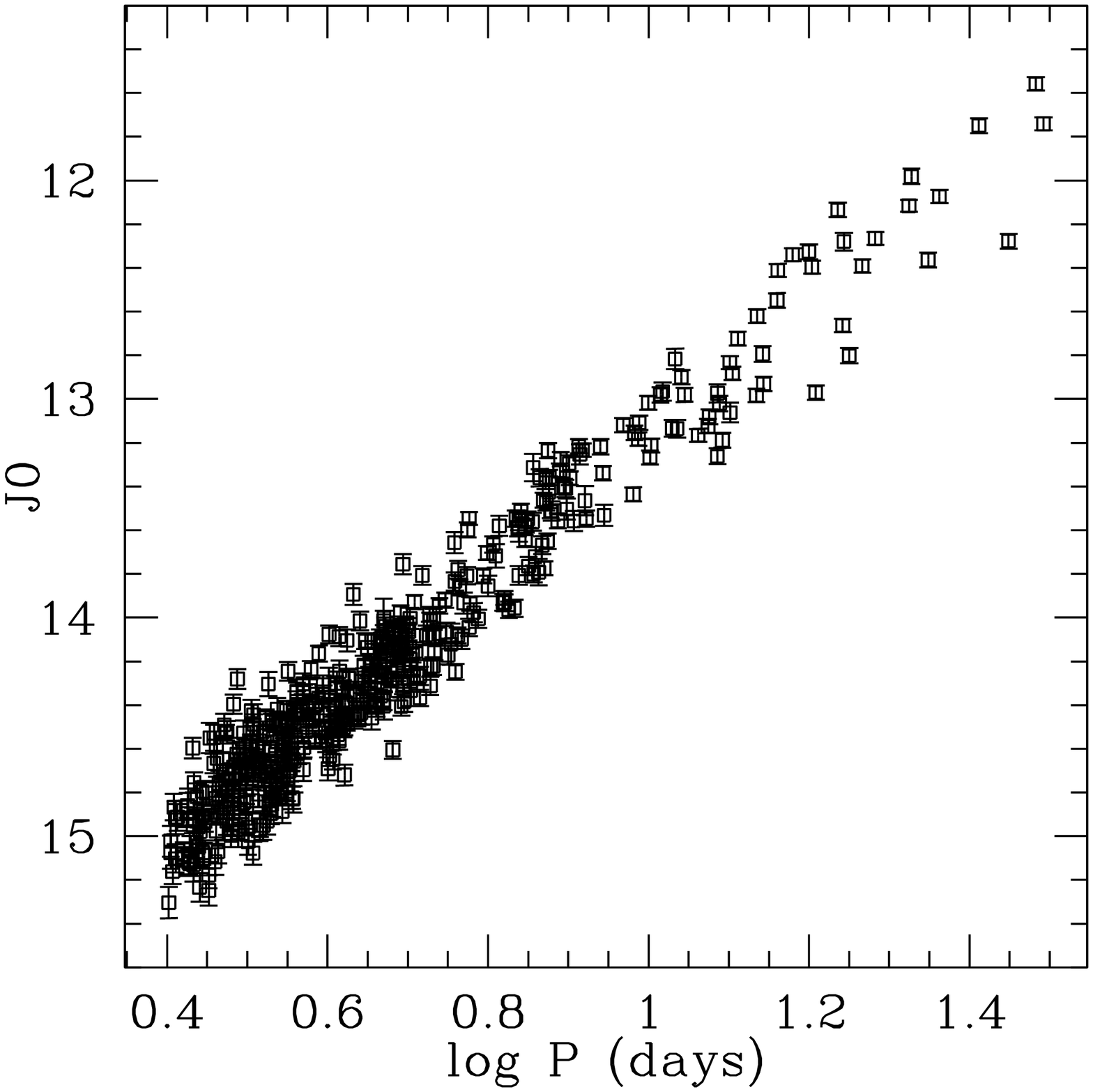}
\includegraphics[width=.4\textwidth]{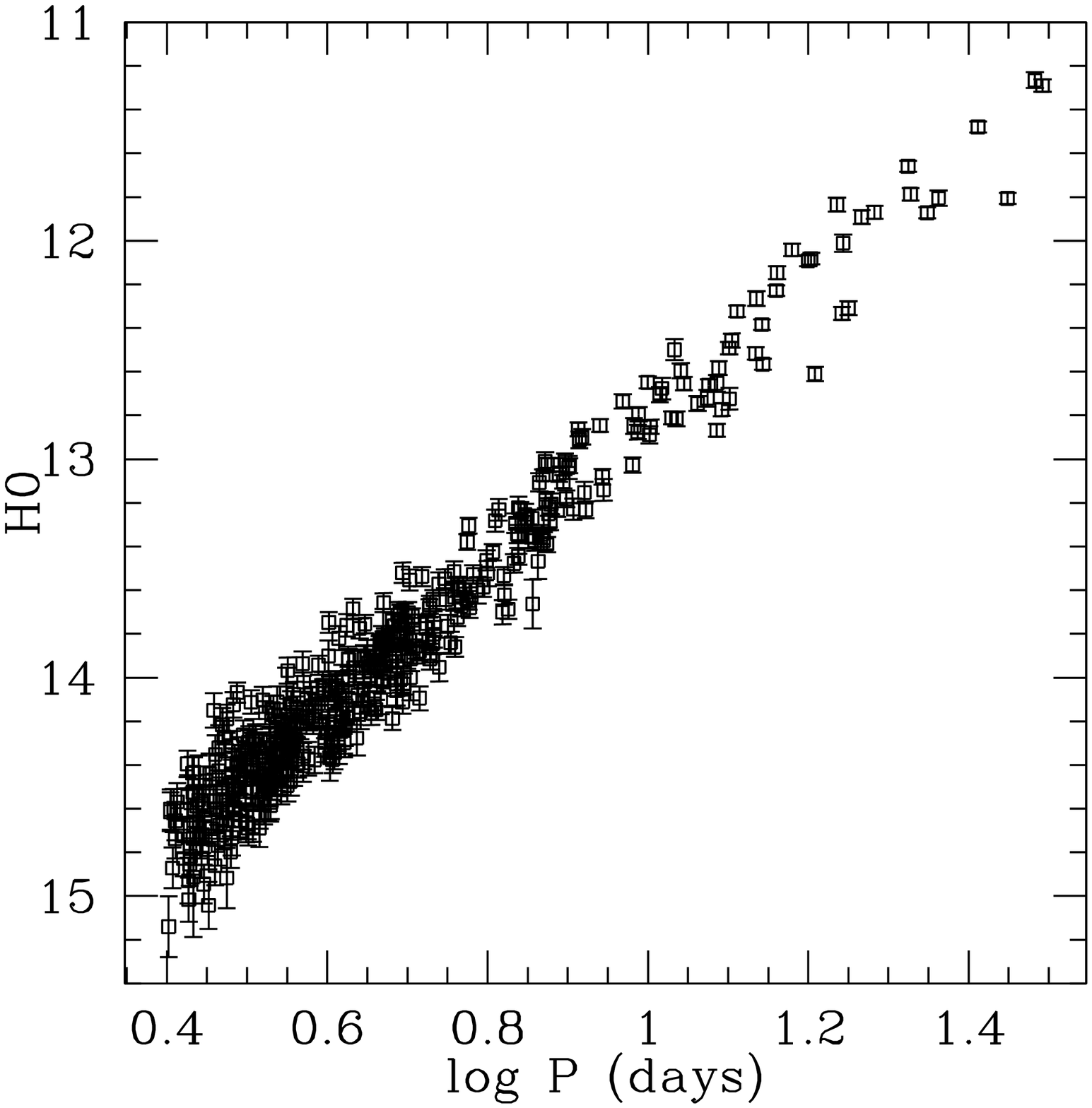}
\includegraphics[width=.4\textwidth]{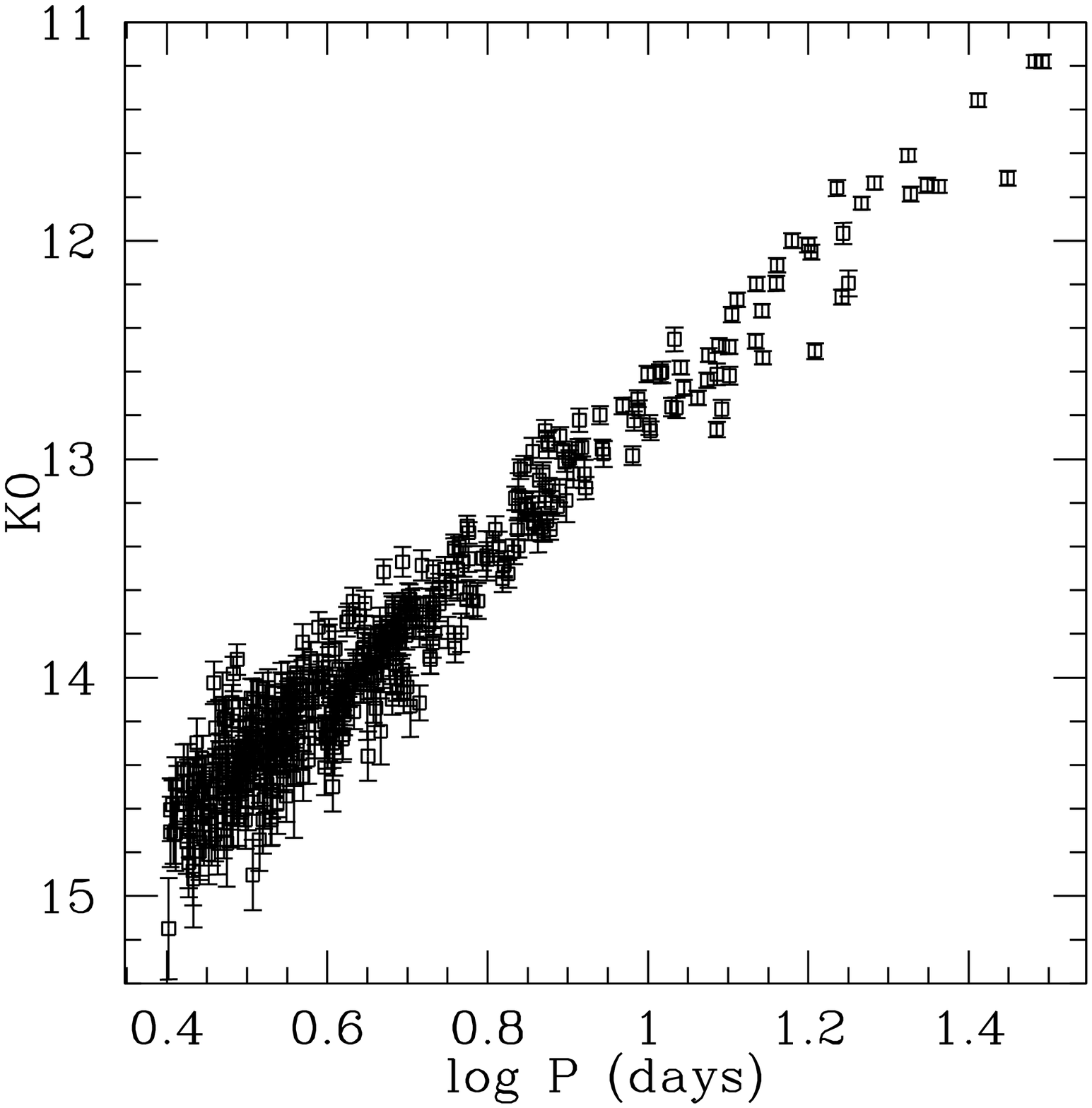}
\end{center}
\caption[]{LMC PL relations in $VIWJHK$ bands. Note that the
relations in the $J$, $H$ and $K$ bands are derived from single-phase data,
which introduces an additional dispersion to the relations not present
in the optical $V$, $I$ and $W$ relations which are based on accurate
mean magnitudes from the OGLE2 database}
\label{Fig3}
\end{figure}

Another possible systematic effect on PL slopes can arise from
differences in the period ranges covered by the LMC and Galactic
Cepheid samples. In $\log P$, it ranges from 0.1 to 1.5 for the LMC
(median 0.59), versus 0.6 to 1.6 for the Milky Way (median
1.16). However, cutting the LMC sample at 0.6 removes more than half
of the OGLE2 sample. We therefore adopted the cut at $\log P = 0.4$,
as done by the OGLE2 team. We also removed a few stars which were
rejected in our linear fits to finally adopt a common sample of 644
stars for $V$, $I$ and $W$, and 447 stars with 2MASS random-phase magnitudes
in $J$, $H$ and $K$. The slopes of the corresponding PL relations derived
from these samples are given in Table~\ref{Tab2} and the LMC PL
relations are shown in Fig.~\ref{Fig3}.

Finally, we tested for the possibility that the different PL slopes
seen in the LMC and Galactic Cepheid samples could actually be an
artifact of our method of distance determination. The most obvious
source which could produce a significant effect on the PL slope of our
Galactic sample is an error in the adopted value of the p-factor used
to convert the Cepheid radial velocity into pulsational velocity. With a
variation of the p-factor within any reasonable limits, however, it is
clearly impossible to recover the slopes, in the different bands, seen
in the LMC Cepheid sample. We also tried to eliminate the most
uncertain distances in our Galactic sample, which happen when there is
an apparent phase shift between the angular and linear diameter
variations in our solutions. Such phase shifts are seen in about one
third of our Galactic Cepheids (always smaller than 5 percent) and are
most likely due to a slight phase mismatch between the radial velocity
curve and the photometric curves used in the analyses, which were not
obtained simultaneously (see a detailed discussion of this in Gieren
et al. (\cite{gfg97})). Eliminating those potentially "problematic"
stars did not change significantly the derived Galactic slopes. Our
adopted distances are based on a bisector linear least-squares fit of
the angular diameters vs. linear displacements at each phase. We also
tested the effect of adopting the inverse fit (all errors assumed to
be carried by the angular diameters) instead of the bisector fit, as
recommmended in Gieren et al. (\cite{gfg97}), without producing
noticeable differences. As a result from these different exercises
carried out on the data, we conclude that our adopted distances are
very robust against these kinds of subtleties.

We therefore adopted two sets of zero points: the first one assumes
our Galactic slopes and the second one assumes the revised LMC
slopes. Results are given in Table~\ref{Tab3}.

\begin{table}
\caption{Absolute magnitudes of a 10-day period Cepheid in $VIWJHK$ bands}
\begin{center}
\renewcommand{\arraystretch}{1.4}
\setlength\tabcolsep{5pt}
\begin{tabular}{lll}
\hline\noalign{\smallskip}
Band & $M_{\rm Galactic \  slopes}$ & $M_{\rm revised \  LMC \  slopes}$ \\
\noalign{\smallskip}
\hline
\noalign{\smallskip}
$B$ & $-3.320 \pm 0.036$ &                    \\
$V$ & $-4.049 \pm 0.034$ & $-4.087 \pm 0.039$ \\
$I$ & $-4.790 \pm 0.034$ & $-4.820 \pm 0.038$ \\
$W$ & $-5.919 \pm 0.032$ & $-5.952 \pm 0.035$ \\
$J$ & $-5.346 \pm 0.029$ & $-5.359 \pm 0.038$ \\
$H$ & $-5.666 \pm 0.031$ & $-5.672 \pm 0.040$ \\
$K$ & $-5.698 \pm 0.031$ & $-5.762 \pm 0.040$ \\
\hline
\end{tabular}
\end{center}
\label{Tab3}
\end{table}

It appears that the choice of slope only very slightly affects the
adopted zero points. This justifies to force the more accurately
determined LMC slopes to our 32 Galactic calibrators, and allows a
determination of the LMC distance in each band. The results are given
in Table~\ref{Tab4}.

\begin{table}
\caption{LMC distance moduli in $VIWJHK$ bands, derived by adopting
the OGLE2 reddenings}
\begin{center}
\renewcommand{\arraystretch}{1.4}
\setlength\tabcolsep{5pt}
\begin{tabular}{ccc}
\hline\noalign{\smallskip}
Band & LMC intercept at 10 days & $\mu_{\rm LMC}$ \\
\noalign{\smallskip}
\hline
\noalign{\smallskip}
$V$ & $14.318 \pm 0.026$ & $18.405 \pm 0.047$ \\
$I$ & $13.631 \pm 0.017$ & $18.451 \pm 0.041$ \\
$W$ & $12.597 \pm 0.009$ & $18.549 \pm 0.036$ \\
$J$ & $13.185 \pm 0.026$ & $18.544 \pm 0.046$ \\
$H$ & $12.853 \pm 0.024$ & $18.525 \pm 0.046$ \\
$K$ & $12.793 \pm 0.026$ & $18.554 \pm 0.048$ \\
\hline
\end{tabular}
\end{center}
\label{Tab4}
\end{table}

The values in Table~\ref{Tab4} show that the distance moduli increase
when the reddening sensitivity of the band decreases.  This is an
annoying result, which we did not see in our 1998 paper (Gieren et
al. \cite{gfg98}).  The main difference is that we then used a reduced
sample of about 60 LMC Cepheids (OGLE results were not available yet),
among which about one half had individual reddening measurements,
which yielded a mean value of $E(B-V)=0.08$, to be compared to the
OGLE2 mean value for Cepheids of 0.147.

We therefore tested the effect of replacing the individual OGLE2
reddenings (which are constant within each of the 84 OGLE2 sub-fields,
but slightly varying from field to field) by a mean value of
$E(B-V)=0.10$, 
as done by the HST Key Project on the Extragalactic Distance
Scale (HST-KP) team. Obviously the zero points of the corresponding PL
relations are modified by this change, and when combined to the
Galactic zero points derived by forcing the new LMC slopes given in
last column of Table~\ref{Tab2} to the Galactic data, we get the LMC
distance moduli shown in Table~\ref{Tab5}. It is clear that the
agreement among the different bands is now much better and, in fact,
quite satisfactory.

\begin{table}
\caption{LMC distance moduli in $VIWJHK$ bands, derived by adopting a
  constant reddening of $E(B-V)=0.10$}
\begin{center}
\renewcommand{\arraystretch}{1.4}
\setlength\tabcolsep{5pt}
\begin{tabular}{ccc}
\hline\noalign{\smallskip}
Band & LMC intercept at 10 days & $\mu_{\rm LMC}$ \\
\noalign{\smallskip}
\hline
\noalign{\smallskip}
$V$ & $14.453 \pm 0.029$ & $18.536 \pm 0.048$ \\
$I$ & $13.713 \pm 0.018$ & $18.530 \pm 0.041$ \\
$W$ & $12.597 \pm 0.009$ & $18.549 \pm 0.036$ \\
$J$ & $13.220 \pm 0.026$ & $18.577 \pm 0.045$ \\
$H$ & $12.873 \pm 0.024$ & $18.544 \pm 0.046$ \\
$K$ & $12.806 \pm 0.026$ & $18.567 \pm 0.048$ \\
\hline
\end{tabular}
\end{center}
\label{Tab5}
\end{table}

We note that the HST-KP for $H_{\circ}$ determination is not fully
consistent in that sense, because they use the OGLE2 PL relations, but
assume at the same time a mean LMC reddening of $E(B-V)=0.10$. If we
use our new LMC PL relations with $E(B-V)=0.10$ (last column of
Table~\ref{Tab2}) and assume a LMC distance modulus of 18.50 as they
did, the resulting zero points are changed and appear in
Table~\ref{Tab6}. It is seen that this introduces a significant
difference in the Cepheid absolute magnitudes in the $V$ and $I$ bands, at
a Cepheid period of 10 days.

To circumvent, or minimize the reddening problem, we prefer to exclude
the $V$ and $I$ band results from our final determination of the distance
modulus to the LMC. In fact, the $W$ value already combines the
information from $V$ and $I$ bands in the best possible way. We therefore
take a weighted mean of the $W$ value on one side and the infrared
weighted average of $J$, $H$, $K$ on the other side, which gives $18.541 \pm
0.047$ for OGLE2 reddenings and $18.563 \pm 0.046$ for a constant
$E(B-V)=0.10$. This gives a greater weight to $W$ which is truly
reddening free, and a lower one to the infrared values which are
derived only from random-phase magnitudes. The uncertainty of the
mean comes from the weighted rms dispersion of the values from all the
bands.

From this procedure we find, as our best adopted value, a LMC
distance modulus of $18.55 \pm 0.06$. The uncertainty does not include
the systematic uncertainty arising if the LMC and Galactic slopes are
really different. In that case, the derived offset depends on the
adopted period for the zero-point. If we measure the offset at the
median value of the LMC sample ($\log P = 0.59$) in place of $\log P =
1$, the derived $W$ modulus becomes 18.41. We are indebted to
Fr\'ed\'eric Pont for this remark.

\subsection{The Hipparcos parallaxes method}

It has been common-place in the past years to present the Galactic
calibration based on Hipparcos parallaxes (\cite{h97}) of about 200
Cepheids as discrepant from other calibrations. This probably arose
from the large distance of the LMC published in the original work of
Feast \& Catchpole (\cite{fc97}), $\mu = 18.70 \pm 0.10$. However, we
will see that the Hipparcos calibration is not discrepant at all, and
that the problem arises in the application of the Hipparcos
calibration to the determination of the LMC distance.

The outstanding idea of Feast \& Catchpole (\cite{fc97}) to combine
the very uncertain, but also very numerous, parallax measurements of
Cepheids by Hipparcos to derive a PL relation zero point for Cepheids
has been shown to be free of biases by subsequent studies (Pont
\cite{p99}, Lanoix et al.  \cite{l+99}, Groenewegen \& Oudmaijer
\cite{go00}). The last of these studies is probably the most accurate
one, and generalizes the result to different photometric bands. We
will adopt their zero points as the Hipparcos Galactic calibration,
based on 236 Cepheids (median $\log P = 0.82$). For details about the
method, the reader is referred to the above references.

For a 10-day period Cepheid, these zero points are given in
Table~\ref{Tab6} and compared to our zero points and to the adopted
zero points of the HST-KP for $H_{\circ}$ determination (Freedman et
al. \cite{f+01}, Macri et al. \cite{m+01}, based on the original OGLE2
LMC relations and on new infrared PL relations, assuming a LMC
distance modulus of 18.50).  Please note that the values of the slopes
and the definitions of $W$ adopted to derive these zero points vary
among these works. For comparison, the original Feast \& Catchpole
(\cite{fc97}) $V$ band zero point was $-4.24 \pm 0.10$. Table~\ref{Tab6}
also gives the HST-KP zero points derived adopting the new OGLE2 LMC
relations based on a mean reddening of $E(B-V)=0.10$.

\begin{table}
\caption{Zero point comparison for a 10-day period Cepheid in $VIWJHK$ bands}
\begin{center}
\renewcommand{\arraystretch}{1.4}
\setlength\tabcolsep{5pt}
\begin{tabular}{ccccc}
\hline\noalign{\smallskip}
Band & $M_{\rm Hipparcos}$ & \multicolumn {2} {c} {$M_{\rm HST-KP}$} & $M_{\rm this \  work}$ \\
\cline{3-4} & & literature & $E(B-V)=0.10$ & \\
\noalign{\smallskip}
\hline
\noalign{\smallskip}
$V$ & $-4.21 \pm 0.11$ & $-4.218 \pm 0.02$ & $-4.047 \pm 0.029$ & $-4.049 \pm 0.034$ \\
$I$ & $-4.93 \pm 0.12$ & $-4.904 \pm 0.01$ & $-4.787 \pm 0.018$ & $-4.790 \pm 0.034$ \\
$W$ & $-5.96 \pm 0.11$ & $-5.899 \pm 0.01$ & $-5.903 \pm 0.009$ & $-5.919 \pm 0.032$ \\
$J$ &                  & $-5.32 \pm 0.06$  & $-5.280 \pm 0.026$ & $-5.346 \pm 0.029$ \\
$H$ &                  & $-5.66 \pm 0.05$  & $-5.627 \pm 0.024$ & $-5.666 \pm 0.031$ \\
$K$ & $-5.76 \pm 0.17$ & $-5.73 \pm 0.05$  & $-5.694 \pm 0.026$ & $-5.698 \pm 0.031$ \\
\hline
\end{tabular}
\end{center}
\label{Tab6}
\end{table}

We must be cautious with the conclusions to be drawn from
Table~\ref{Tab6}: there is an apparently good agreement between the
Hipparcos and the original HST-KP zero points on one hand, and between
the ISB Galactic and the revised HST-KP zero points on the other
hand. How is this to be interpreted?

First of all, if the Hipparcos and the original HST-KP zero points
agree, why do they lead to different distance moduli for the LMC?
Simply because the adopted LMC PL relations have different intercepts:
Feast \& Catchpole (\cite{fc97}) adopted a LMC PL relation in $V$ from
Caldwell \& Laney (\cite{cl91}) based on 88 Cepheids with an intercept
of $14.42 \pm 0.02$, and based on a mean adopted reddening of
$E(B-V)=0.08$ 
(30 have individual $BVI$ reddenings), while Freedman et
al. (\cite{f+01}) used the originally published OGLE2 PL relations
based on more than 600 Cepheids with an intercept of $14.282 \pm
0.021$ and a mean reddening of 0.147.  The observed difference of 0.14
mag in intercepts is well explained by the difference in adopted mean
reddenings ($0.067 \times 3.3 = 0.22$) and is sufficient to make the
distance moduli discrepant. Please note that Feast \& Catchpole
(\cite{fc97}) also added a metallicity correction of 0.04 mag, even
increasing the discrepancy.

\begin{figure}[ht]
\begin{center}
\includegraphics[width=.5\textwidth]{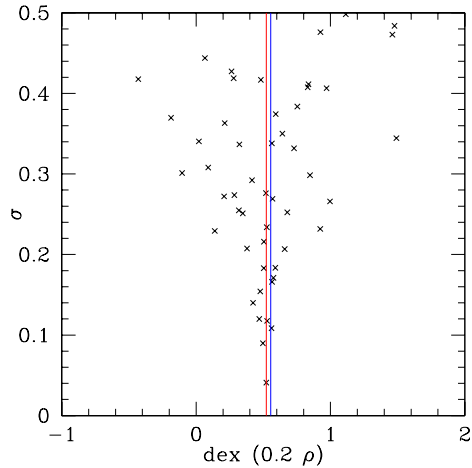}
\end{center}
\caption[]{Hipparcos $V$ PL relation individual zero points (expressed
as $10^{\, 0.2 \, \rho}$, where $\rho$ is the zero point at $\log P =
1$) vs. their uncertainty; superimposed as vertical lines are the
Hipparcos adopted mean (red) and the alternative ISB value (blue)}
\label{Fig4}
\end{figure}

In fact, the low accuracy of the Hipparcos zero points makes the
observed difference between the Hipparcos and the ISB $V$ zero points
not significant, as can be seen in Fig.~\ref{Fig4} which displays, for
the Cepheids with the highest weights, the value of the zero point
estimate $10^{\, 0.2 \, \rho}$ vs. its uncertainty, together with the
positions of the adopted Hipparcos and the ISB zero points. In the $I$
band, the zero point difference is smaller, and clearly not
significant if we consider the alternative value published by Lanoix
et al. (\cite{l+99}), which is $-4.86 \pm 0.09$. Finally, there is a
good agreement in $W$- and $K$-band zero points, but this is probably
quite fortuitous for the same reasons.

Now, the excellent agreement between the revised HST-KP and the
Galactic ISB Cepheid absolute magnitudes at a 10-day period is clearly
more significant thanks to the high accuracy of both results. This
basically demonstrates that the HST-KP adopted LMC distance modulus of
18.50 is very nearly correct.

\subsection{The ZAMS-fitting calibration}

Feast (\cite{f199}) published a revised list of 31 Galactic Cepheids
belonging to open clusters or associations with distance moduli
derived through the ZAMS-fitting technique. He also explained why the
values may not be modified by the Pleiades distance change after the
Hipparcos measurement (simultaneous change of the reference ZAMS of
the same order of magnitude).  In fact, his cluster distance values
are very close to those published in Gieren \& Fouqu\'e
(\cite{gf93}).

Twenty-four of these cluster Cepheids lie in the same period range of
our Galactic calibrators of the ISB technique. We have derived PL
relations from these Cepheids, which are given below, and appear to be
in good agreement with those derived from the ISB distances, although
they are less accurate. They support the evidence that the Galactic PL
slopes are somewhat steeper than the LMC ones.

\begin{eqnarray}
M_v & = & -2.767 \pm 0.173 \  (\log P - 1) - 4.160 \pm 0.055 \  (\sigma = 0.271 \  N=24) \\
M_i & = & -3.273 \pm 0.164 \  (\log P - 1) - 4.837 \pm 0.051 \  (\sigma = 0.218 \  N=18) \\
M_w & = & -3.684 \pm 0.144 \  (\log P - 1) - 5.980 \pm 0.045 \  (\sigma = 0.191 \  N=18) \\
M_k & = & -3.766 \pm 0.170 \  (\log P - 1) - 5.694 \pm 0.051 \  (\sigma = 0.217 \  N=18)
\end{eqnarray}

Very recently, Turner \& Burke (\cite{tb02}) published a revised list
of 46 Cepheids belonging to clusters or associations. Fifteen of these
Cepheids have ISB distances in our current sample. The weighted mean
of the distance moduli differences we find is not significant and
amounts to:

\begin{equation}
\left< \mu (\rm ISB) - \mu (\rm ZAMS) \right> = +0.01 \pm 0.06 \  \sigma = 0.24
\end{equation}

\noindent
after rejection of AQ Pup ($0.79 \pm 0.11$ difference - we note that
cluster membership of this star seems to be very uncertain). Other
large differences are observed for $\delta$ Cep ($0.37 \pm 0.11$), BB
Sgr ($0.49 \pm 0.08$), and U Car ($-0.45 \pm 0.05$).  Excluding these
stars, for which the case of membership in their respective
clusters/associations is not strong (see original references cited in
the paper of Turner \& Burke), the rms dispersion is 0.10,
corresponding to 3\% distance precision for each set.

From this comparison, we conclude that the Cepheid distance scale
based on ZAMS-fitting is consistent with the calibration from the ISB
method. This may be a bit surprising given the many difficulties in
the application of the ZAMS-fitting method, and the doubts shed on the
method after Hipparcos.

\section{What may still be wrong in the LMC distance?}

\subsection{Reddening effects}

We have seen previously how changing the adopted reddening may change
our results. By reddening we mean both the reddening values and the
reddening law. There is good evidence in the literature that the LMC
reddening law may differ from the Galactic one shortward of $B$ (see,
e.g., Gochermann \& Schmidt-Kaler \cite{gs02}). However, this is of
little concern for us. What is more important is that there is some
evidence that the $R_v$ value may be lower in the LMC. For instance,
Misselt et al. (\cite {m+99}) find $R_v$ values varying between 2.16
and 3.31, and obtain a good fit using the standard Cardelli et
al. (\cite{c+89}) reddening law for a mean value of $R_v =
2.4$. Similar, but slightly higher values (between 2.66 and 3.60) have
been found in the SMC by Gordon \& Clayton (\cite {gc98}). So, what we
interpret as a smaller mean LMC reddening than the OGLE2 values may in
fact be due to a lower $R_v$ value.

Concerning the reddening values, several studies have investigated
both the foreground reddening due to our Galaxy (the LMC is at
$-33^{\circ}$ galactic latitude), and the internal reddening. They 
show that the reddening is patchy, with large variations from a line
of sight to another.

Concerning the foreground reddening, Schwering \& Israel (\cite{si91})
find from 48' resolution maps a range of 0.06 to 0.17 in $E(B-V)$, with
an average value of 0.10.  By comparison, the foreground reddening in
front of the SMC is found more homogeneous, only varying from 0.06 to
0.08. In the LMC, Oestreicher et al. (\cite{o+95}) with a better
resolution of 10' also find a large range from 0 to 0.15, with an
average value of $0.06 \pm 0.02$.

Concerning the internal reddening, Oestreicher \& Schmidt-Kaler
(\cite{os96}) find a range of 0.06 to 0.29, with a mean $E(B-V)=0.16$,
while Harris et al. (\cite{h+97}) find a total average extinction of
0.20, from which they conclude that the mean internal reddening
amounts to $E(B-V)=0.13$ mag.

In comparison, Udalski et al. (\cite{u+99}) use the mean magnitude of
the Red Giant Clump (RGC) to derive the total reddening variations along the
21 LMC OGLE2 fields (mainly along the bar).  They divide each field
into 4 sub-fields and give a mean reddening along each of the 84 lines
of sight, corresponding to a resolution of 14.2'. The zero point of
their reddening scale is given by three photometric measurements from
the literature. They find a range of total extinction between 0.105
and 0.201, with a mean value over the fields of 0.137, and a mean
Cepheid value of $E(B-V)=0.147$.

However, Girardi \& Salaris (\cite {gs01}) have shown that the RGC
mean absolute magnitude depends on population effects (age and
metallicity). This implies that the OGLE2 method is only valid as long
as it can be assumed that the population characteristics do not change
along the LMC bar.

\begin{figure}[ht]
\begin{center}
\includegraphics[width=.7\textwidth]{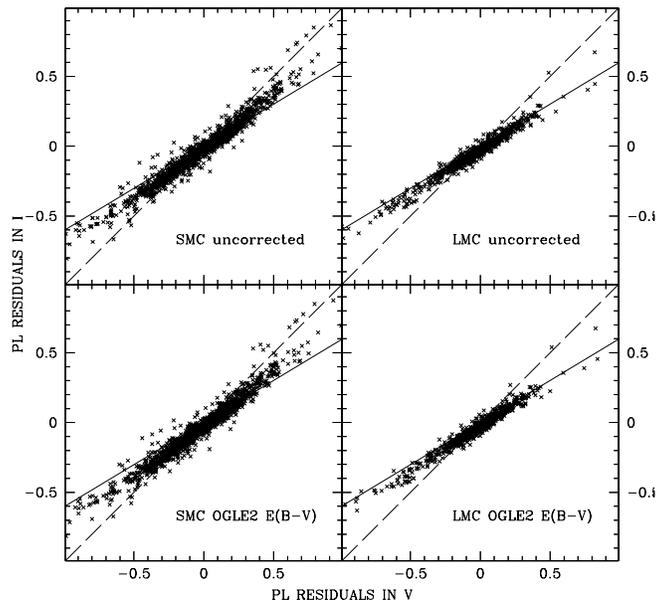}
\end{center}
\caption[]{Plot of PL relations residuals in $V$ and $I$ from Beaulieu et
  al. (\cite{b+01}). The data from the LMC clearly correlate along the
  reddening line (solid) even after application of the OGLE2 reddening
  correction}
\label{Fig5}
\end{figure}

In any case, Beaulieu et al. (\cite{b+01}) have shown that the
resolution of the OGLE2 maps is not sufficient to consider their
reddenings as individual values for each Cepheid, since the PL
residuals in $V$ and $I$ correlate along the reddening line in the case of
LMC, as can be seen from Fig.~\ref{Fig5}, reproduced from their paper.

It is therefore tempting to use $BVI$ OGLE2 measurements to derive
individual reddenings, following the Dean et al. (\cite{d+78})
precepts, adapted to the LMC metallicity as in Caldwell \& Coulson
(\cite{cc85}). Such measurements exist for 329 Cepheids, which is
about one half of the calibrating sample. Unfortunately, the result
is disappointing, because when we apply the derived individual
reddenings to correct the mean $V$ and $I$ magnitudes, the dispersions
of the PL relations {\it increase}.
 
Finally, some authors use period-colour (PC) relations to estimate the
reddenings. This is the case for instance in the various works based
on Hipparcos parallaxes. However, it is well known that the PC
relations have considerable intrinsic dispersion, but as shown by
Feast \& Catchpole (\cite{fc97}), these over- or under-estimated
reddenings compensate in part for the intrinsic width of the PL
relations.  But, as the Cepheid colours vary with metallicity at a
given period (see below), it is important to use a PC relation adapted
to the sample under study. Only 7 Hipparcos Cepheids do not have reddenings
measurements in Fernie's database (\cite{f+95}). We have therefore
checked that using these individual reddenings in place of those
derived from a PC relation indeed gives a similar result for the Galactic
zero point in $V$.

As a summary from this discussion, it is clear that the reddening
problem is still far from being overcome, but we believe that our
current adopted procedures do minimize the influence of reddening on
the LMC distance modulus derived in the previous section. Once more,
the merits of using infrared bands or reddening insensitive parameters
such as $W$ are underlined.

\subsection{Metallicity effects}

\subsubsection{Theoretical point of view}

There is a long debate in the literature on the effect metallicity may
have on the PL relations in different photometric bands, both on the
theoretical and observational sides.  From an observer's point of
view, it seems that one can always find a theory which agrees with the
metallicity dependence one finds by observational tests. However, not
all the theories rest on the same basis. It is well known that purely
radiative stellar pulsation models predict too large pulsation
amplitudes for Cepheids. Some convective transport must be added, for
instance by means of the Mixing Length Theory (MLT). However,
time-independent MLT cannot predict the position of the red edge of
the instability strip, which needs the additional introduction of a
time-dependent dissipation introduced by the eddy viscosity. But
time-dependent MLT models are not successful if they are too local in
space. We therefore need at least a non-local time-dependent
hydrodynamic pulsation code to correctly describe the coupling of
pulsation and convection.

There are not so many codes available. Some linearize the equations
(Yecko et al. \cite{y+98}) while others solve the full non-linear
equations (Bono \& Stellingwerf \cite{bs94}, Feuchtinger
\cite{f299})). It seems that their results concerning metallicity
dependences basically agree.  The fact that they disagree with
predictions of models based on purely radiative pulsation codes which
neglect the coupling of pulsation with convection (Saio \& Gautschy
\cite{sg98}, Alibert et al. \cite{a+99}) seems to us an effect of
these simplifications. We will therefore basically follow the
predictions of the full-amplitude (non-linear) models including a
non-local and time-dependent treatment of stellar convection from Bono
et al. (\cite{b1+99}). But we are aware that these models depend on a
number of not well constrained parameters, the adopted values of which
may change the predictions to a significant level (see Figs.~12 and 13
in Yecko et al. \cite{y+98}).

These models first predict that at a given metallicity (Y and Z
fixed), the width of the instability strip changes from low- to
high-mass Cepheids, therefore invalidating older model predictions,
which assumed that the red edge was parallel to the blue edge. Now,
for a given Cepheid mass (and therefore luminosity), an increase in Z
(and Y) shifts the instability strip towards cooler effective
temperatures, due to a decrease in the pulsation destabilization
caused by the hydrogen ionization region; therefore, at a fixed period
and {\it assuming} a uniformly populated instability strip, metal-poor
pulsators are {\it more} luminous than metal-rich ones.

However, this prediction for bolometric magnitudes does not
necessarily apply to all photometric bands, and differences in the
adopted atmosphere models may generate differences in the magnitude of
the effect for a given band. Bono et al. (\cite{b2+99}) find that the
dependence on metallicity is increased in the $V$ band, due to the
dependence of the bolometric correction on effective temperature,
while it is smaller in the $K$ band. But the effect is still that
metal-poor pulsators are more luminous than metal-rich ones, when
absolute magnitudes are derived from a PL relation. However,
metallicity also affects colours, and an increase in Z at fixed period
gives redder $B-V$ and $V-K$ colours. When using a PLC relation to derive
the absolute magnitudes, both effects must be taken into account, and
at fixed period {\it and} colour, metal-poor pulsators are still slightly
{\it more} luminous in $K$ but {\it less} luminous in $V$ than metal-rich ones.

This theoretical model also predicts that the slopes of the PL
relation vary with metallicity, in the sense that an increase in Z
produces shallower slopes in the $V$ and $K$ bands, the effect being
smaller in $K$. We observe the opposite effect. Also, the slopes of the
PC relations ($B-V$ and $V-K$) are predicted to steepen when Z increases.

\subsubsection{Observational point of view}

On the observational side, let us start with differential studies. By
this, we mean studies in different regions of a given galaxy showing a
spatial variation in metallicity, to assess magnitude differences, at
a given period, due to the variation of metallicity. The pioneering
work of Freedman \& Madore (\cite{fm90}) compared Cepheids in three
fields of M~31 at different galactocentric distances; it led to
unconclusive results, varying from no significant metallicity effect
according to the original authors to large effects according to
Gould's re-analysis (\cite{g94}). A more accurate study was conducted
in M~101 by Kennicutt et al. (\cite{k+98}) who compared the derived
distance moduli for 2 fields, located at different radial distances
from the center of the galaxy having a difference of 0.7 dex in
nebular [O/H]. They found some effect in the sense that distances to
metal-rich galaxies are underestimated when derived by using the LMC
Cepheid PL relation, but the share of effect between $V$ and $I$ bands is
not specified.

In the same spirit, we have observed the Sculptor Group spiral NGC~300
in $B$, $V$ and $I$ and discovered about 120 Cepheids, well distributed all
over the galaxy (Pietrzy\'nski et al. \cite{p+02}). By measuring the
stellar metallicity in different regions from B and A supergiants
spectra, we plan to measure any differential effect due to
metallicity. This work is in progress, and we believe that this will
provide the as yet most stringent observational test on the
metallicity sensitivity of the Cepheid PL relation. We have also
observed outer disc Cepheids of our own galaxy (Pont et
al. \cite{p+01}), with the hope that the metallicity difference to the
solar region would mimic the metallicity difference to the LMC. It
appears, however, that the metallicity range only reaches the typical
LMC metallicity ([Fe/H] $\sim -0.3$) at about 14 kpc, where very few
Cepheids are known (Luck et al. \cite{lu+03}). This makes evidencing
metallicity effects in our own galaxy a difficult task. The task could seem
easier when comparing Galactic to SMC Cepheids, as recently shown by
Storm et al. (\cite{s+02}), but here again disentangling metallicity
effects from uncertain reddenings for SMC stars, depth and ridge
line effects for such a small sample (5 stars) is quite a challenge.

Another approach was pioneered by Beaulieu et al. (\cite{b+97}) and
Sasselov et al. (\cite{s+97}) in the Magellanic Clouds for $V$ and $I$
bands, and generalized by Kochanek (\cite{k97}) to 17 galaxies in
$UBVRIJHK$ bands.  From an analysis of 481 Cepheids detected by the EROS
microlensing experiment in the LMC and SMC, and assuming that the
slopes of the PL relations do not depend on metallicity, Beaulieu and
Sasselov found that an SMC Cepheid is {\it less} luminous than a LMC
Cepheid of same period by 0.06 mag in the blue EROS band (intermediate
between Johnson $B$ and $V$) but {\it more} luminous by 0.01 mag in the
red EROS band (intermediate between Cousins $R$ and $I$). The net effect
is to overestimate the SMC distance modulus by 0.14 mag.  Kochanek
finds that metal-poor pulsators are {\it more} luminous than
metal-rich ones in $U$ and $B$, but {\it less} luminous in $VIJHK$, with the
difference increasing with the wavelength. These studies then
translate the metallicity dependence into a distance modulus variation
per dex of [Fe/H], but nothing proves so far that such a dependence is
linear.

Finally, Udalski et al. (\cite{u+01} presented recently an analysis of
PL relations in IC~1613, a galaxy of even lower metallicity than the
SMC ([Fe/H] $\sim -1.0$), and showed that the slopes of the $V$ and $I$
relations were not significantly different from those found in the
LMC, giving a strong observational hint that the slopes do not depend
on metallicity, at least in the range $-1.0$ to $-0.3$ in [Fe/H]. They
also argue that the zero points do not depend on metallicity, but this
relies on comparison with other distance indicators which themselves
depend upon metallicity, so this result currently lies on less stable
grounds.

From all these theoretical and observational results on the
metallicity effect on Cepheid absolute magnitudes currently available,
it is our impression that if a metallicity dependence of the PL
relations exists, it should be small, its sign is currently not well
defined and may depend on the photometric band, and it may not be a
linear function of [Fe/H].

\section{Consequence for other distance indicators}

At the time of this writing, no distance indicator can claim to be so
accurate that other distance indicators become unnecessary. All
distance indicators suffer, to some extent, from systematic
uncertainties and the best way to constrain the Extragalactic Distance
Scale seems to compare the results of various distance indicators for
which previous work has shown that they provide relatively accurate
measures of distances.

It is not the purpose of this review to compare in detail the
calibrations of all the most promising distance indicators. Other
review papers in this book deal with them. We just want to find out
what our preferred Cepheid PL relation derived in this paper predicts
for several other of the most common distance indicators.

For this purpose, we only need to know the difference in magnitude
between a Cepheid of 10-day pulsation period and the following other
distance indicators: Tip of the Red Giant Branch magnitude (TRGB), Red
Giant Clump mean magnitude (RGC), and RR Lyrae magnitude. We adopt the
following values of differences from Udalski (\cite{u00}), based on
several nearby galaxies and his adopted corrections for different
metallicities:

\begin{eqnarray}
V_{\circ} \  ({\rm RR \  Lyrae \  at \  [Fe/H]=-1.6}) - V_{\circ} \  ({\rm Cepheid \  at \  10 \  days}) & = & 4.60 \\
(V-I)_{\circ} \  ({\rm Cepheid \  at \  10 \  days}) & = & 0.70 \\
I_{\circ} \  ({\rm TRGB}) - I_{\circ} \  ({\rm Cepheid \  at \  10 \  days}) & = & 0.70 \\
I_{\circ} \  ({\rm RGC \  at \  [Fe/H]=-0.5}) - I_{\circ} \  ({\rm TRGB}) & = & 3.60
\end{eqnarray}

From these differences and the Galactic ISB zero points from
Table~\ref{Tab3}, it is easy to predict expected values for other
distance indicators. Our current Cepheid calibration corresponds to:

\begin{eqnarray}
M_v \  ({\rm RR \  Lyrae \  at \  [Fe/H]=-1.6}) & = & +0.55 \\
M_i \  ({\rm TRGB}) & = & -4.09 \\
M_i \  ({\rm RGC \  at \  [Fe/H]=-0.5}) & = & -0.49 
\end{eqnarray}

We leave to others the discussion of the importance of population
effects on these differences to determine which precise value should
be applied in any case, and the comparison of our predicted values to
the range of published values in the literature. We just want to note
the very good agreement with the RR Lyrae mean $V$ magnitude presented
at this conference by C. Cacciari and G. Clementini, namely $M_v =
+0.59 \pm 0.03$ at $\rm [Fe/H] = -1.5$, which corresponds to $+0.57$
at $\rm [Fe/H] = -1.6$.

\section{Conclusions}

We have used the infrared surface brightness technique to obtain a new
absolute calibration of the Cepheid PL relation in optical and
near-infrared bands from improved data on Galactic stars. The infrared
surface brightness distances to the Galactic variables are consistent
with direct interferometric Cepheid distance measurements, and with
the PL calibration coming from Hipparcos parallaxes of nearby
Cepheids, but are more accurate than these determinations. We find
that in all bands, the Galactic Cepheid PL relation appears to be
slightly, but significantly steeper than the corresponding relation
defined by the LMC Cepheids. This systematic difference has recently
been confirmed by Tammann et al. (\cite{t+03}) and could be a
signature of a metallicity effect on the slope of the PL
relation. Since the slope of our LMC Cepheid sample is clearly better
defined than the one of the much smaller Galactic sample, we fit the
LMC slopes to our Galactic calibrating Cepheid sample (which
introduces only a small uncertainty) to obtain our final, adopted and
improved absolute calibrations of the Cepheid PL relations in the
$VIWJHK$ bands. Comparing the absolute magnitudes of 10-day period
Cepheids in both galaxies which are only slightly affected by the
different Galactic and LMC slopes of the PL relation, we derive values
for the LMC distance modulus in all these bands which can be made to
agree extremely well under reasonable assumptions for both, the
reddening law, and the adopted reddenings of the LMC
Cepheids. However, reddening remains an important and not
satisfactorily resolved issue, and in order to obtain a LMC distance
determination as independent of reddening as possible, we adopt as our
final result a weighted mean of the values coming from the
reddening-insensitive Wesenheit magnitude, and those derived from the
near-infrared bands. This yields, as our current best estimate from
Cepheid variables, a LMC distance modulus of $18.55 \pm 0.06$.

A discussion of the effect of metallicity on Cepheid absolute
magnitudes as provided by both, existing empirical and theoretical
evidence makes us conclude that at the present time, it seems likely
that there is some metallicity dependence of the PL relation, of small
size, whose sign is not clear, and whose size may depend on the
photometric band. It may also be a non-linear function of metallicity,
with some indication that the metallicity effect on the Cepheid PL
relation does not change basically between very low and LMC
metallicities, but that the slope of the metallicity dependence may
steepen when going from LMC to solar abundances. Clearly, more work
from both theory and, particularly, from the observational side has to
be done to improve the constraints on the metallicity effect.  Until
this is achieved, it may be the best choice to use our current,
Galactic calibration in applications to the distance measurement of
Cepheids in solar metallicity galaxies. Thanks to its accuracy
provided by the infrared surface brightness technique, the Galactic
calibration is now a true alternative to using the LMC calibration,
with the added benefit of minimizing metallicity-related effects when
studying Cepheid samples in metal-rich spiral galaxies.

\section{Acknowledgements}

The authors are indebted to Jean-Philippe Beaulieu for comments on the
validity of theoretical predictions concerning the metallicity effects
on the PL relations and for authorizing us to reproduce Figure~\ref{Fig5}.
Giuseppe Bono also helped us to better understand the current
differences among theories.

Some calculations would not have been possible without Martin
Groenewegen's kindness, who sent his list of $JHK$ magnitudes for OGLE2
Cepheids and basic data for Hipparcos-observed Cepheids.

Comments and corrections from Fr\'ed\'eric Pont and Daniel Cordier are
very gratefully acknowledged, as well as valuable inputs from Sergei
Andrievsky, David Bersier, Pierre Kervella, Francesco Kienzle, Lucas
Macri, Barbara Mochej\-ska, Georges Paturel, Eric Persson, and
Grzegorz Pietrzy\'nski.

WG gratefully acknowledges financial support for this work from the
Chilean Center for Astrophysics FONDAP 15010003.

\begin{table}[t]
\caption{Data for the 32 Galactic calibrators, from our new infrared
  surface brightness analysis of these stars. $R$ is the star mean
  radius in solar units, and $\sigma_R$ its uncertainty}
\begin{center}
\renewcommand{\arraystretch}{1.4}
\setlength\tabcolsep{2pt}
\begin{tabular}{lcrlrlcccccccc}
\hline\noalign{\smallskip}
%\scriptsize
ID & $\log P$ & $\mu_{\circ}$ & $\sigma_{\mu}$ & $R$ & $\sigma_R$ & $M_B$ & $M_V$ & $M_I$ & $M_J$ & $M_H$ & $M_K$ & $M_W$ & $E(B-V)$ \\
\noalign{\smallskip}
\hline
\noalign{\smallskip}
BF Oph       & 0.609329 &  9.271 & 0.034 &  32.0 &  0.5 & $-2.13$ & $-2.75$ & $-3.40$ & $-3.84$ & $-4.11$ & $-4.18$ & $-4.37$ &  0.247 \\
T Vel        & 0.666501 &  9.802 & 0.060 &  33.6 &  0.9 & $-2.05$ & $-2.69$ & $-3.37$ & $-3.88$ & $-4.18$ & $-4.26$ & $-4.39$ &  0.281 \\
$\delta$ Cep & 0.729678 &  7.084 & 0.044 &  42.0 &  0.9 & $-2.87$ & $-3.43$ & $-4.06$ & $-4.47$ & $-4.75$ & $-4.81$ & $-5.01$ &  0.092 \\
CV Mon       & 0.730685 & 10.988 & 0.034 &  40.3 &  0.6 & $-2.46$ & $-3.04$ & $-3.80$ & $-4.26$ & $-4.54$ & $-4.65$ & $-4.93$ &  0.714 \\
V Cen        & 0.739882 &  9.175 & 0.063 &  42.0 &  1.2 & $-2.71$ & $-3.30$ & $-3.96$ & $-4.41$ & $-4.69$ & $-4.77$ & $-4.95$ &  0.289 \\
BB Sgr       & 0.821971 &  9.519 & 0.028 &  49.8 &  0.6 & $-2.82$ & $-3.52$ & $-4.26$ & $-4.72$ & $-5.02$ & $-5.10$ & $-5.38$ &  0.284 \\
U Sgr        & 0.828997 &  8.871 & 0.022 &  48.4 &  0.5 & $-2.82$ & $-3.51$ & $-4.25$ & $-4.70$ & $-4.98$ & $-5.06$ & $-5.35$ &  0.403 \\
$\eta$ Aql   & 0.855930 &  6.986 & 0.052 &  48.1 &  1.1 & $-2.94$ & $-3.58$ & $-4.27$ & $-4.71$ & $-5.01$ & $-5.07$ & $-5.31$ &  0.149 \\
S Nor        & 0.989194 &  9.908 & 0.032 &  70.7 &  1.0 & $-3.34$ & $-4.10$ & $-4.86$ & $-5.41$ & $-5.73$ & $-5.82$ & $-6.00$ &  0.189 \\
Z Lac        & 1.036854 & 11.637 & 0.055 &  77.8 &  2.0 & $-3.86$ & $-4.56$ & $-5.29$ & $-5.71$ & $-6.02$ & $-6.09$ & $-6.40$ &  0.404 \\
XX Cen       & 1.039548 & 11.116 & 0.023 &  69.5 &  0.7 & $-3.43$ & $-4.16$ & $-4.90$ & $-5.42$ & $-5.72$ & $-5.80$ & $-6.02$ &  0.260 \\
V340 Nor     & 1.052579 & 11.145 & 0.185 &  67.1 &  5.7 & $-2.98$ & $-3.82$ & $-4.68$ & $-5.22$ & $-5.58$ & $-5.67$ & $-5.98$ &  0.315 \\
UU Mus       & 1.065819 & 12.589 & 0.084 &  74.0 &  2.9 & $-3.42$ & $-4.16$ & $-4.92$ & $-5.50$ & $-5.81$ & $-5.90$ & $-6.08$ &  0.413 \\
U Nor        & 1.101875 & 10.716 & 0.060 &  76.3 &  2.1 & $-3.71$ & $-4.42$ & $-5.14$ & $-5.65$ & $-5.92$ & $-6.02$ & $-6.23$ &  0.892 \\
BN Pup       & 1.135867 & 12.950 & 0.050 &  83.2 &  1.9 & $-3.76$ & $-4.51$ & $-5.27$ & $-5.78$ & $-6.10$ & $-6.18$ & $-6.40$ &  0.438 \\
LS Pup       & 1.150646 & 13.556 & 0.056 &  90.2 &  2.3 & $-3.93$ & $-4.69$ & $-5.43$ & $-5.96$ & $-6.28$ & $-6.36$ & $-6.56$ &  0.478 \\
VW Cen       & 1.177138 & 12.803 & 0.039 &  86.6 &  1.5 & $-3.15$ & $-4.04$ & $-4.93$ & $-5.63$ & $-6.02$ & $-6.13$ & $-6.28$ &  0.448 \\
X Cyg        & 1.214482 & 10.421 & 0.016 & 105.3 &  0.8 & $-4.12$ & $-4.99$ & $-5.77$ & $-6.28$ & $-6.62$ & $-6.69$ & $-6.94$ &  0.288 \\
VY Car       & 1.276818 & 11.501 & 0.022 & 112.9 &  1.1 & $-3.93$ & $-4.85$ & $-5.70$ & $-6.33$ & $-6.68$ & $-6.78$ & $-7.00$ &  0.243 \\
RY Sco       & 1.307927 & 10.516 & 0.034 & 100.0 &  1.5 & $-4.40$ & $-5.06$ & $-5.81$ & $-6.27$ & $-6.54$ & $-6.62$ & $-6.93$ &  0.777 \\
RZ Vel       & 1.309564 & 11.020 & 0.029 & 114.7 &  1.5 & $-4.25$ & $-5.04$ & $-5.82$ & $-6.40$ & $-6.73$ & $-6.82$ & $-7.00$ &  0.335 \\
WZ Sgr       & 1.339443 & 11.287 & 0.047 & 121.8 &  2.6 & $-3.87$ & $-4.80$ & $-5.72$ & $-6.38$ & $-6.76$ & $-6.88$ & $-7.10$ &  0.467 \\
WZ Car       & 1.361977 & 12.918 & 0.066 & 112.0 &  3.4 & $-4.14$ & $-4.92$ & $-5.72$ & $-6.32$ & $-6.66$ & $-6.74$ & $-6.92$ &  0.384 \\
VZ Pup       & 1.364945 & 13.083 & 0.057 &  97.1 &  2.5 & $-4.32$ & $-5.01$ & $-5.72$ & $-6.19$ & $-6.49$ & $-6.56$ & $-6.79$ &  0.471 \\
SW Vel       & 1.370016 & 11.998 & 0.025 & 117.5 &  1.4 & $-4.21$ & $-5.02$ & $-5.85$ & $-6.44$ & $-6.79$ & $-6.89$ & $-7.09$ &  0.349 \\
T Mon        & 1.431915 & 10.777 & 0.053 & 146.3 &  3.6 & $-4.36$ & $-5.33$ & $-6.21$ & $-6.85$ & $-7.24$ & $-7.34$ & $-7.53$ &  0.209 \\
RY Vel       & 1.449158 & 12.019 & 0.032 & 139.9 &  2.1 & $-4.69$ & $-5.50$ & $-6.30$ & $-6.88$ & $-7.18$ & $-7.28$ & $-7.51$ &  0.562 \\
AQ Pup       & 1.478624 & 12.522 & 0.045 & 147.9 &  3.1 & $-4.65$ & $-5.51$ & $-6.41$ & $-6.95$ & $-7.30$ & $-7.40$ & $-7.75$ &  0.512 \\
KN Cen       & 1.531857 & 13.124 & 0.045 & 185.8 &  3.9 & $-5.64$ & $-6.33$ & $-6.98$ & $-7.50$ & $-7.83$ & $-7.94$ & $-7.94$ &  0.926 \\
l Car        & 1.550855 &  8.989 & 0.032 & 201.7 &  3.0 & $-4.71$ & $-5.82$ & $-6.77$ & $-7.45$ & $-7.87$ & $-7.96$ & $-8.20$ &  0.170 \\
U Car        & 1.589083 & 10.972 & 0.032 & 161.5 &  2.4 & $-4.72$ & $-5.62$ & $-6.48$ & $-7.10$ & $-7.45$ & $-7.56$ & $-7.78$ &  0.283 \\
RS Pup       & 1.617420 & 11.622 & 0.076 & 214.7 &  7.5 & $-5.11$ & $-6.08$ & $-7.02$ & $-7.66$ & $-8.03$ & $-8.14$ & $-8.45$ &  0.446 \\
\hline
\end{tabular}
\end{center}
\label{Tab7}
\end{table}

%INDEX%%%%%%%%%%%%%%%%%%%%%%%%%%%%%%%%%%%%%%%%%%%%%%%%%%%%%%%%%%%%%%%
% Please check with the editor of your book whether he plans to
% include a "mutual" subject index - if so, please code your entries
% in the standard syntax. For your own purposes you may print your
% "personal" index by using the following commands:
%
%\clearpage
%\addcontentsline{toc}{section}{Index}
%\flushbottom
%\printindex
%%%%%%%%%%%%%%%%%%%%%%%%%%%%%%%%%%%%%%%%%%%%%%%%%%%%%%%%%%%%%%%%%%%%%

\end{document}